\documentstyle[emulateapj,psfig]{article}

\newbox\grsign \setbox\grsign=\hbox{$>$}
\newdimen\grdimen \grdimen=\ht\grsign
\newbox\laxbox \newbox\gaxbox
\setbox\gaxbox=\hbox{\raise.5ex\hbox{$>$}\llap
     {\lower.5ex\hbox{$\sim$}}}\ht1=\grdimen\dp1=0pt
\setbox\laxbox=\hbox{\raise.5ex\hbox{$<$}\llap
     {\lower.5ex\hbox{$\sim$}}}\ht2=\grdimen\dp2=0pt
\newcommand{\gax}{\mathrel{\copy\gaxbox}}
\newcommand{\lax}{\mathrel{\copy\laxbox}}

\begin{document}

\submitted{To appear in ApJ 570, May 10 2002}

\title{Extended Power-Law Decays in BATSE Gamma-Ray Bursts: \\
         Signatures of External Shocks?}

\author{T. W. Giblin\altaffilmark{1,2,7}, V. Connaughton\altaffilmark{1,2},
J. van Paradijs\altaffilmark{1,3,6}, R. D. Preece\altaffilmark{1,2}, 
M. S. Briggs\altaffilmark{1,2}, C. Kouveliotou\altaffilmark{2,4},
R.A.M.J. Wijers\altaffilmark{5}, G. J. Fishman\altaffilmark{2}}

\altaffiltext{1}{Department of Physics, University of Alabama in Huntsville,
Huntsville, AL 35899, USA}
\altaffiltext{2}{NASA Marshall Space Flight Center, SD50, Huntsville, AL
35812, USA}
\altaffiltext{3}{Astronomical Institute ``Anton Pannekoek'', University
of Amsterdam, \& Center for High Energy Astrophysics, Kruislaan 403,
1098 SJ Amsterdam, The Netherlands}
\altaffiltext{4}{Universities Space Research Association}
\altaffiltext{5}{Department of Physics and Astronomy, SUNY, Stony Brook,
NY 11794-3800, USA}
\altaffiltext{6}{Deceased}
\altaffiltext{7}{New address:  Department of Physics and Astronomy, The 
College of Charleston, 58 Coming St., Charleston, SC 29424, USA}

\begin{abstract}

The connection between Gamma-Ray Bursts (GRBs) and their afterglows is 
currently not well understood. Afterglow models of synchrotron emission 
generated by external shocks in the GRB fireball model predict emission 
detectable in the gamma-ray regime ($\gax 25$ keV). In this paper, we 
present a temporal and spectral analysis of a subset of BATSE GRBs with 
smooth extended emission tails to search for signatures of the ``early 
high-energy afterglow'', i.e., afterglow emission that initially begins 
in the gamma-ray phase and subsequently evolves into X-Ray, uv, optical, 
and radio emission as the blast wave is decelerated by the ambient medium. 
From a sample of 40 GRBs we find that the temporal decays are best 
described with a power-law $\sim t^{\beta}$, rather than an exponential, 
with a mean index $\langle \beta \rangle \approx -2$. Spectral analysis
shows that $\sim 20\%$ of these events are consistent with fast-cooling
synchrotron emission for an adiabatic blast wave; three of which are 
consistent with the blast wave evolution of a jet, with $F_{\nu} \sim t^{-p}$. 
This behavior suggests that, in some cases, the emission may originate from a 
narrow jet, possibly consisting of ``nuggets'' whose angular size are less 
than $1 / \Gamma$, where $\Gamma$ is the bulk Lorentz factor.

\end{abstract}


\section{Introduction}

Afterglow emissions from Gamma-Ray Bursts (GRBs) in the X-ray, 
optical, and radio wavebands are in good agreement with afterglow 
models of relativistic fireballs (\cite{wijers97}, \cite{galama98}, 
\cite{waxman97}, \cite{vietri97}). The observed afterglow spectrum 
is well-described as synchrotron emission that arises from the
interaction of the relativistic blast wave with bulk Lorentz factor 
$\Gamma_{0} \sim 10^2 - 10^3$ with the ambient medium (\cite{meszaros97},
\cite{galama98}). The often highly variable gamma-ray phase of the 
burst may reflect the physical behavior of the fireball progenitor 
through collisions internal to the flow, i.e., internal shocks 
(\cite{sari97}, \cite{kobayashi97}). On the other hand, Dermer 
and Mitman (1999) have suggested a blast wave with an inhomogenous 
external medium. Heinz and Begelman (1999) have suggested an 
inhomogeneous bullet-like jet outflow that encounters the 
interstellar medium.
 
The precise relationship between the observed GRB and the afterglow 
emission is not well understood. GRBs recorded by the {\it BeppoSAX} 
satellite suggest that the X-ray afterglow emission may be delayed in 
time from the main GRB (e.g., GRB970228, \cite{costa00}) or may begin 
during the GRB emission (e.g., GRB980519, \cite{intzand99}). In the 
latter case, it is not clear if the X-ray afterglow is a separate 
underlying emission component or a continuation of the GRB itself. 
The internal-external shock model presents a scenario in which emission 
from internal and external shocks may overlap in time. If the internal 
shocks reflect the activity of the progenitor, then the onset of the 
afterglow may be separated from the prompt gamma-ray emission. Since 
the nature of the progenitor is not known, the effect of the ambient 
medium on the emission from the progenitor is highly problematic. The 
model therefore does not prohibit internal and external shock emissions 
from overlap, while in other cases the afterglow emission may be delayed 
with respect to the GRB (e.g., \cite{sari99a}, \cite{meszaros99}, 
\cite{vietri00}). 

The detection of optical emission simultaneous with the gamma-ray 
emission of GRB990123 (\cite{akerlof99}) provided the first evidence
for two distinct emission components in a GRB; here, the prompt optical 
emission is believed to originate from synchrotron emission in the
production of the reverse shock generated when the ejecta encounters 
the external medium (\cite{galama99}, \cite{sari99a}, \cite{meszaros99}). 
The gamma-ray spectrum of GRB990123 can not be extrapolated from the 
spectral flux from the simultaneous optical emission, indicating that 
the optical and gamma-ray emission originate from two separate 
mechanisms (\cite{briggs99}, \cite{galama99}). Evidence for overlapping 
shock emission was also found in GRB980923 (\cite{giblin99a}), where a 
long power-law decay tail ($\sim t^{-1.8}$) was observed in soft
gamma-rays (25-300 keV). Two separate emission components are favored
in this burst because the spectral characteristics of the tail were 
markedly different from those of the variable main GRB emission. The 
spectrum in the tail is consistent with that of a slow-cooling synchrotron 
spectrum, similar to the behavior of low-energy afterglows (e.g., 
\cite{bloom98}, \cite{vreeswijk99}).

The gamma-rays produced by internal shocks and the soft gamma-rays of 
the ``afterglow'' may therefore overlap, the latter having a signature 
of power-law decay in the synchrotron afterglow model. If this is the 
case, at least some GRBs in the BATSE database should show signatures 
of the early external shock emission. These events would contain a soft 
gamma-ray (or hard X-ray) tail component that decays as a power-law in 
their time histories, possibly superposed upon the variable gamma-ray 
emission. It has been shown that the peak frequency of the initial 
synchrotron emission, which depends on the parameters of the system 
(see $\S 2$), can peak in hard X-rays or gamma-rays (\cite{meszaros92}). 
Further, it may be possible to see a smoothly decaying GRB that is the 
result of an external shock, i.e., the GRB itself is a ``high-energy'' 
afterglow. For such GRBs, the subsequent afterglow emission in X-rays 
and optical would then simply be the evolution of the burst spectrum. 
A situation like this might arise when the progenitor generates only 
a single energy release (i.e., no internal shocks). 

It is well known that the temporal structures of GRBs are very diverse 
and often contain complex, rapid variability. However, some bursts 
exhibit smooth decay features that persist on timescales as long as, 
or even longer than, the variable emission of the burst. Our 
investigation focuses on the combined temporal and spectral behavior 
of a sample of 40 BATSE GRBs that exhibit smooth decays during the 
later phase of their time histories. Many of these events fall into 
a category of bursts traditionally referred to as ``FREDs'' (Fast 
Rise, Exponential-like Decay), bursts with rapid rise times and a 
smooth extended decay (\cite{kouveliotou92}). In $\S 2$ we present 
temporal and spectral properties of the afterglow synchrotron spectrum. 
In $\S 3$ we examine the temporal behavior and spectral characterisitics 
of the decay emission for the events in our sample and compare their 
spectra with the model synchrotron spectrum. A color-color diagram 
(CCD) technique is also applied to systematically explore the spectral 
evolution of each event. In $\S 4$ we present a set of high-energy 
afterglow candidates, followed by a discussion of our results in the 
framework of current fireball models.

\section{Synchrotron Spectra from External Shocks}

Internal shocks are capable of liberating some fraction of the 
total fireball energy $E_{0} = \Gamma_{0} M_{0} c^{2}$, leaving 
a significant fraction to be injected into the external medium 
via the external shock (\cite{kobayashi97}). However, recent
simulations suggest that internal shock efficiencies can approach 
$\sim 100\%$ (\cite{beloborodov00}). Nonetheless, as the blast 
wave sweeps up the external medium, it produces a relativistic 
forward shock and a mildly relativistic reverse shock in the 
opposite direction of the initial flow. The reverse shock 
decelerates the ejecta while the forward shock continuously 
accelerates the electrons into a non-thermal distribution of 
energies described by a power law  
$dn_{e}/d\gamma_{e} \propto \gamma_{e}^{-p}$, where $\gamma_{e}$ 
is the electron Lorentz factor. The distribution has a low-energy 
cutoff given by $\gamma_{m} \le \gamma_{e}$. Behind the shock, the 
accelerated electrons and magnetic field acquire some fraction, 
$\epsilon_{e}$ and $\epsilon_{B}$ of the internal energy. 

The resulting synchrotron spectrum of the relativistic electrons 
consists of four power-law regions (\cite{sari98}) defined by three 
critical frequencies $\nu_{\rm a}$, $\nu_{\rm c}$, and $\nu_{\rm m}$, 
where $\nu_{\rm a}$ is the self-absorption frequency, 
$\nu_{\rm c} = \nu(\gamma_{\rm c})$ is the cooling frequency, and 
$\nu_{\rm m} = \nu(\gamma_{\rm m})$ is the characteristic synchrotron 
frequency (see Figure 1 in \cite{sari98}). Here, we are only concerned 
with the high-energy spectrum, therefore we do not consider self-absorption. 
Electrons with $\gamma_{e} \ge \gamma_{\rm c}$ cool down to $\gamma_{\rm c}$, 
the Lorentz factor of an electron that cools on the hydrodynamic timescale of 
the shock (\cite{piran99}). The electrons cool rapidly when 
$\gamma_{\rm m} \ge \gamma_{\rm c}$, known as {\it fast-cooling} 
(i.e., $\nu_{\rm m} > \nu_{\rm c}$), and cool more slowly when 
$\gamma_{\rm m} \le \gamma_{\rm c}$, known as {\it slow-cooling}. 
In the fast-cooling regime, the evolution of the shock may range 
from fully radiative ($\epsilon_{e} \sim 1$) to fully adiabatic 
($\epsilon_{e} \ll 1$). In the slow-cooling mode, the evolution 
can only be adiabatic, since $\gamma_{\rm m} < \gamma_{\rm c}$. 
The characteristic synchrotron frequency of an electron with 
minimum Lorentz factor $\gamma_{\rm m}$ is (\cite{sari99a})
\begin{equation}
   \nu_{\rm m} = 1.0 \times 10^{19}\ {\rm Hz} 
   {\epsilon_{e} \overwithdelims () 0.1}^{2}
   {\epsilon_{B} \overwithdelims () 0.1}^{1/2}
   {\Gamma \overwithdelims () 300}^{4} n_{1}^{1/2},
\end{equation}
corresponding to a break in the observed spectrum with energy
\begin{equation}
   E_{\rm m} = 41.4\ {\rm keV} {\epsilon_{e} \overwithdelims () 0.1}^{2}
   {\epsilon_{B} \overwithdelims () 0.1}^{1/2}
   {\Gamma \overwithdelims () 300}^{4} n_{1}^{1/2},
\end{equation}
where $\Gamma$ is the bulk Lorentz factor and $n_{1}$ is the constant 
density of the ambient medium. Although the frequency in equation 1 
depends strongly on the parameters of the system, the forward shock 
may very well peak initially in hard X-rays or in gamma rays (Sari 
and Piran 1999). 

The synchrotron spectrum evolves with time according to the 
hydrodynamic evolution of the shock and the geometry of the 
fireball (e.g., spherical or collimated). Specifically, the 
time dependence of $\nu_{\rm c}$ and $\nu_{\rm m}$ will strongly 
depend on the time evolution of the Lorentz factor 
$\gamma (t)$. Assuming a spherical blast wave and a homogeneous
medium, for radiative fast-cooling, $\nu_{\rm m} \propto t^{-12/7}$ 
and $\nu_{\rm c} \propto t^{-2/7}$, while for adiabatic evolution 
(fast or slow-cooling) $\nu_{\rm m} \propto t^{-3/2}$ and 
$\nu_{\rm c} \propto t^{-1/2}$. The shape of the synchrotron 
spectrum remains constant with time as $\nu_{\rm c}$ and $\nu_{\rm m}$ 
evolve to lower values. In the fast-cooling mode, $\nu_{\rm m}$ 
decays faster than $\nu_{\rm c}$, causing a transition in the 
spectrum from fast to slow-cooling. 

Since the break frequencies scale with time as a power-law, the 
spectral energy flux of the synchrotron spectrum, $F_{\nu}$ (erg 
s$^{-1}$ cm$^{-2}$ keV$^{-1}$), will also scale as a power-law in 
time so that $F_{\nu}(\nu,t) \propto \nu^{\alpha}t^{\beta}$, where 
the spectral and temporal power-law indices, $\alpha$ and $\beta$, 
depend on the temporal ordering of $\nu_{\rm c}$ relative to 
$\nu_{\rm m}$, i.e., fast or slow-cooling. For radiative 
fast-cooling,
\begin{equation}
   F_{\nu} \propto 
   \cases{
      \nu^{1/3} t^{-1/3}         & $\nu < \nu_{\rm c}$,\cr
      \nu^{-1/2} t^{-4/7}        & $\nu_{\rm c} < \nu < \nu_{\rm m}$,\cr
      \nu^{-p/2} t^{(2-6p)/7}    & $\nu_{\rm m} < \nu$,\cr
         }
\end{equation}
and for adiabatic fast-cooling,
\begin{equation}
   F_{\nu} \propto 
   \cases{
      \nu^{1/3} t^{1/6}          & $\nu < \nu_{\rm c}$,\cr
      \nu^{-1/2} t^{-1/4}        & $\nu_{\rm c} < \nu < \nu_{\rm m}$,\cr
      \nu^{-p/2} t^{(2-3p)/4}    & $\nu_{\rm m} < \nu$\cr
         }
\end{equation}
(\cite{sari98}). For slow-cooling the spectral energy flux is
\begin{equation}
   F_{\nu} \propto
   \cases{
      \nu^{1/3} t^{1/2}             & $\nu < \nu_{\rm m}$,\cr
      \nu^{-(p-1)/2} t^{-3(p-1)/4}  & $\nu_{\rm m} < \nu < \nu_{\rm c}$,\cr
      \nu^{-p/2} t^{-(3p-2)/4}      & $\nu_{\rm c} < \nu$\cr
         }
\end{equation}
(\cite{sari98}). Note that a simple relation exists between the 
temporal and spectral indices through the value of the electron 
index $p$ for the high-energy spectral slopes ($\nu > \nu_{\rm c}$) 
and the spectral slope below $\nu_{\rm c}$ in the slow-cooling regime. 
Defining the low-energy spectral slope as $\alpha$ and the high-energy 
spectral slope as $\alpha^{\prime}$, the following relations between 
the temporal and spectral indices for a spherical blast wave are 
established (\cite{sari98}):
\begin{equation}
   \beta =
      \cases{
      2(6 \alpha^{\prime} + 1)/7 & ({\rm fast-cooling, radiative}),\cr
      3 \alpha^{\prime}/2 + 1/2  & ({\rm fast-cooling, adiabatic}),\cr
      3 \alpha/2 &({\rm slow-cooling}, $\nu_{\rm m} < \nu < \nu_{\rm c}$),\cr
      3 \alpha^{\prime}/2 + 1/2 &({\rm slow-cooling}, $\nu_{\rm c} < \nu$).\cr
      }
\end{equation}
The numerical value of $p$ is readily determined from the measured 
high-energy spectral slope, $p = -2 \alpha^{\prime}$. Long wavelength 
afterglow measurements give typical electron indices in the range 
$2.0 \le p \le 2.5$. Although the nature of the emission in this model
is always synchrotron radiation with its characteristic slopes and breaks, 
the time dependence of the breaks are affected by the details of the 
geometry and dynamics.

The relations in equation 6 are only valid in the case of a spherical 
blast wave encountering a constant density medium. Rhoads (1999) 
considered the adiabatic evolution of a collimated or jet-like outflow 
in which the ejecta are confined to a conical volume with a half opening 
angle $\theta_{c}$. As the outflow encounters the external medium, the 
bulk Lorentz factor of the flow, $\Gamma$, decreases with radius and time 
as a power-law (e.g., see \cite{huang99}). However, the hydrodynamical 
evolution of the shock changes from a power-law to an exponential regime 
when $\theta_{b} \equiv \Gamma^{-1} \simeq \theta_{c}$ (\cite{rhoads99}, 
\cite{sari99b}). The observer is able to discern that the flow is confined 
to an expanding cone rather than a sphere because less radiation is observed. 
In consequence, a break in the light curve to a $F_{\nu} \sim t^{-p}$ 
behavior is observed as the ejecta sweep up a larger amount of mass. 
For adiabatic evolution of a jet, $\nu_{\rm m} \propto t^{-2}$, 
$\nu_{\rm c} \propto t^{0} =$ const, and the peak flux scales as
$F_{\nu,{\rm max}} \propto t^{-1}$ (\cite{rhoads99}, \cite{sari99b}). 
Thus the spectral flux for an adiabatic jet in the fast-cooling regime 
is given by
\begin{equation}
   F_{\nu} \propto 
   \cases{
      \nu^{1/3} t^{-1}          & $\nu < \nu_{\rm c}$,\cr
      \nu^{-1/2} t^{-1}         & $\nu_{\rm c} < \nu < \nu_{\rm m}$,\cr
      \nu^{-p/2} t^{-p}         & $\nu > \nu_{\rm m},$\cr
         }
\end{equation}
and for slow-cooling,
\begin{equation}
   F_{\nu} \propto
   \cases{
      \nu^{1/3} t^{-1/3}          & $\nu < \nu_{\rm m}$,\cr
      \nu^{-(p-1)/2} t^{-p}       & $\nu_{\rm m} < \nu < \nu_{\rm c}$,\cr
      \nu^{-p/2} t^{-p}           & $\nu > \nu_{\rm c}.$\cr
         }
\end{equation}
The jet geometry can therefore be tested by the simple relation 
$\beta = 2\alpha^{\prime} = -p$, irrespective of whether the spectrum 
is fast or slow-cooling, provided that 
$\nu > {\rm max}(\nu_{\rm c},\nu_{\rm m})$.

\section{Analysis}

We examine the properties of extended decay emission in GRBs in the energy 
range $\sim 25$-2000 keV using data from BATSE, a multi-detector all-sky 
monitor instrument onboard the {\it Compton Gamma-Ray Observatory} (CGRO). 
BATSE consisted of eight identical detector modules placed at the corners 
of the CGRO in the form of an octahedron (\cite{fishman89}). Each module 
contains a Large Area Detector (LAD) composed of a sodium iodide crystal 
scintillator that continuously recorded count rates in 1.024 and 2.048 
second time intervals with four and sixteen energy channels, respectively 
(known as the DISCLA and CONT data types). Nominally, a burst trigger is 
declared when the count rates in two or more LADs exceed the background 
count rate by at least $5.5\sigma$. Various burst data types are then 
accumulated, including the four channel high time resolution (64 ms) 
discriminator science data (DISCSC). The DISCSC and DISCLA rates cover 
four broad energy channels in the 25-2000 keV range (25-50, 50-100, 100-300, 
$> 300$ keV). The CONT data span roughly the same energy range, but with 
sixteen energy channels and 2.048 s time resolution.

\subsection{Dataset and Background Modeling}

Our dataset was collected by visually selecting events from the 
current BATSE catalog with extended decay features, using DISCSC time 
histories in the 25-2000 keV range. Time histories used in this search 
had a time resolution of 64 ms or longer, therefore our scan was not 
sensitive to the selection of events from the short class of bursts in 
the bimodal duration distribution (\cite{kouveliotou93a}). A study of 
decay emission in short GRBs will not be included in this analysis but 
will be the subject of future work. Our search resulted in a sample of 
40 bursts, 17 with a FRED-like profile and 23 that exhibit a period of 
variability followed by a smooth decaying emission tail. 

We grouped events into three categories based on the characteristic 
time history of the bursts: (1) pure FREDs (PF), (2) FREDs with initial 
variability mainly during the peak (FV), and (3) bursts with a period 
of variability followed by an emission tail (V+T). Note that this 
categorization only serves as a descriptive guideline for this analysis 
and does not imply a robust temporal classification scheme. Our analysis 
uses discriminator (DISCSC and DISCLA) and continuous (CONT) data from 
the BATSE LADs. 

The source count rates in the $i{\rm th}$ time bin and the $j{\rm th}$
energy channel, $S_{i,j}$, were obtained by subtracting the background 
model rates, $B_{i,j}$, from the burst time history. The background model 
rates in the $j{\rm th}$ energy channel were generated by modeling pre 
and post-burst background intervals appropriate for each burst with a 
polynomial of order $n$, where $1 \le n \le 4$. Post-burst intervals 
were chosen at sufficiently late times beyond the tail of the burst, 
since the time when the tail emission disappears into the background 
is somewhat uncertain. This method was adequate for bursts with 
durations less than $\sim 200$ seconds. For longer bursts however, 
the long term variations in the background can inhibit knowledge 
of when the tail emission drops below the background level. For this 
reason, we applied an orbital background subtraction method to events 
with durations that exceed $\sim 200$ seconds. This technique uses as 
background the average of the CONT data count rates registered 
when CGRO's orbital position is at the point closest in geomagnetic 
latitude to that at the time of the burst on days before and after 
the burst trigger. A complete description of the technique is given 
in Connaughton (2000).

\subsection{Temporal Modeling}

In the context of afterglow models, the decay emission is usually fit 
with a power-law. We fit the smooth decay of the background subtracted 
source count rates, $S_{i}$, in each burst with a power-law function of 
the form 
\begin{equation}
	R(t_{i}) = R_{0}(t_{i}-t_{0})^{\beta},  
\end{equation}
where $R(t_{i})$ is the model count rate of the $i{\rm th}$ time bin. 
The free parameters of the model are the amplitude, $R_{0}$, in counts 
s$^{-1}$, the power-law index, $\beta$, and the fiducial point of 
divergence, $t_{0}$, given in seconds. As a general guideline, the fit 
intervals $[\tau_{1},\tau_{2}]$ were selected in a systematic manner. 
For the PF bursts, the start time of each fit interval, $\tau_{1}$, was 
taken as the time of half-width at half-maximum intensity (HWHM) of the 
burst. This approach obviously does not apply to the V+T group of bursts. 
For these events, $\tau_{1}$ was defined as the bin following the apparent 
end time of the variable emission. The fit interval end time, $\tau_{2}$, 
was defined as the time when the amplitude of the tail count rates first 
falls within $1 \sigma_{b,i}$ of the background model, where $\sigma_{b,i}$ 
is the Poisson count rate uncertainty of the $i{\rm th}$ bin of the 
background model. To avoid obtaining a premature value of $\tau_{2}$ 
caused by statistical fluctuations in the tail, the amplitude of the 
count rates in the tail was calculated using a moving average of 16 
time bins. The value of $\tau_{2}$ was not particularly sensitive to 
the width of the moving average. We further found that the fitted model 
parameter values were generally insensitive to arbitrarily larger values 
of $\tau_{2}$.

\begin{deluxetable}{lcclcccc}
\footnotesize
\tablecaption{Summary of Temporal Fits (25-300 keV) \label{tbl-1}}
\tablewidth{0pt}
\tablehead{
\colhead{GRB} & \colhead{Trigger} & 
\colhead{Time Profile\tablenotemark{(1)}} & 
\colhead{$[\tau_{1},\tau_{2}]$ (s)} &
\colhead{$\beta$} & \colhead{$t_{0}$} & \colhead{$d.o.f.$} & 
\colhead{$\chi^{2}/d.o.f.$} 
}
\startdata
910602 &257 &FV &[19,361] &$-1.74_{+0.11}^{-0.72}$ &$-13.69_{+4.41}^{-22.58}$ 
&166 &1.41 \\
910814c &676 &V+T &[67,109] &$-1.75_{+0.40}^{-0.65}$ &$58.05_{+2.05}^{-3.07}$ 
&39 &0.98 \\
910927 &829 &FV &[15,52] &$-2.06_{+0.44}^{-0.72}$ &$7.69_{+2.05}^{-2.14}$ 
&34 &1.20 \\
911016 &907 &FV &[140,292] &$<-2.06$ &$<113.73$ &147 &4.03 \\
920218 &1419 &FV &[130,204] &$-2.15_{+0.33}^{-0.54}$ &$120.90_{+2.08}^{-1.02}$ 
&71 &1.32 \\
920502 &1578 &FV &[14,99] &$-2.51_{+0.22}^{-3.27}$ &$4.69_{+1.10}^{-17.71}$ 
&81 &1.04 \\
920622 &1663 &V+T &[21,80] &$<-2.98$ &$<4.77$ &56 &1.75 \\
920801 &1733 &PF &[9,123] &$-1.91_{+0.33}^{-0.56}$ &$-2.84_{+3.06}^{-2.49}$ 
&110 &2.15 \\
920813 &1807 &FV &[39,150] &$-2.10_{+0.04}^{-6.05}$ &$18.07_{+0.47}^{-118.1}$ 
&106 &1.57 \\
920901 &1885 &PF &[19,200] &$<-2.02$ &$<-12.81$ &175 &1.15 \\
921207 &2083 &FV &[12,56] &$-2.88_{+0.02}^{-0.03}$ &$3.74_{+0.02}^{-0.03}$ 
&41 &1.32 \\
930106 &2122 &FV &[28,503] &$-1.76_{+0.12}^{-0.33}$ &$-12.28_{+5.08}^{-9.23}$ 
&213 &0.88 \\                                                                   
930131 &2151 &V+T &[3,96] &$-0.71_{+0.11}^{-0.31}$ &$-2.41_{+2.12}^{-4.76}$ 
&89 &0.91 \\
930612 &2387 &PF &[17,264] &$-2.09_{+0.11}^{-0.86}$ &$1.01_{+1.48}^{-9.46}$ 
&190 &1.24 \\
931223 &2706 &PF &[10,72] &$-2.57_{+0.10}^{-6.71}$ &$-4.75_{+1.03}^{-60.4}$ &59 
&1.26\\
940218 &2833 &FV &[7,50] &$-3.28_{+0.04}^{-0.08}$ &$-0.83_{+0.03}^{-0.05}$ &40 
&1.60 \\
940419b &2939 &PF &[17,264] &$<-1.75$ &$<-17.48$ &239 &1.51 \\
941026 &3257 &PF &[17,131] &$<-1.95$ &$<-8.89$ &109 &1.67 \\
951104 &3893 &V+T &[22,99] &$-1.97_{+0.02}^{-0.22}$ &$14.49_{+1.03}^{-0.05}$ 
&75 &1.08 \\
960530(1)\tablenotemark{(2)} &5478 &PF &[9,121] &$-1.49_{+0.04}^{-0.77}$ 
&$2.36_{+0.29}^{-6.38}$ &108 &0.88 \\
960530(2)\tablenotemark{(3)} &5478 &PF &[273,515] &$<-2.13$ &$<251.46$ &234 &1.33 \\
970302 &6111 &PF &[9,133] &$-1.49_{+0.11}^{-0.93}$ &$-0.81_{+1.15}^{-10.43}$ &119 
&1.85 \\
970411 &6168 &FV &[44,398] &$-2.06_{+0.20}^{-0.21}$ &$16.01_{+4.23}^{-0.00}$ 
&171 &0.95 \\
970925 &6397 &PF &[12,85] &$-1.98_{+0.21}^{-0.31}$ &$-1.65_{+2.25}^{-1.02}$ &69 
&1.71 \\
971127 &6504 &PF &[11,198] &$-1.96_{+0.11}^{-0.61}$ &$-4.26_{+1.57}^{-6.43}$ 
&180 &1.44 \\
971208 &6526 &PF\tablenotemark{(4)} &[361,2995] &$-1.34_{+0.01}^{-0.11}$ 
&$-26.34_{+6.55}^{-7.05}$ &1047 &2.32 \\
980301 &6621 &PF &[35,87] &$-2.50_{+0.05}^{-1.58}$ &$27.65_{+0.11}^{-7.20}$ 
&48 &1.37 \\
980306 &6629 &FV &[242,402] &$-1.71_{+0.50}^{-0.71}$ &$219.03_{+9.51}^{-5.46}$ 
&154 &1.09 \\
980325 &6657 &PF &[20,152] &$-2.32_{+0.04}^{-5.64}$ &$-7.49_{+1.58}^{-102.5}$ 
&127 &2.36 \\
980329 &6665 &V+T &[16,59] &$<-2.39$ &$<6.64$ &40 &2.48 \\
981203 &7247 &FV &[60,851] &$-1.61_{+0.002}^{-0.013}$ &$-17.1_{+2.13}^{-6.15}$ 
&384 &1.52 \\
981205 &7250 &FV &[18,127] &$-1.59_{+1.20}^{-1.57}$ &$-10.92_{+27.72}^{-14.67}$ 
&104 &0.90 \\
990102 &7293 &PF &[13,235] &$-2.14_{+0.13}^{-1.36}$ &$-3.7_{+2.0}^{-15.71}$ 
&215 &1.14 \\
990220 &7403 &PF &[17,196] &$-1.97_{+0.30}^{-1.04}$ &$-9.78_{+6.16}^{-16.86}$ 
&85 &0.83 \\
990316 &7475 &PF &[25,158] &$<-2.20$ &$<11.06$ &128 &1.54 \\
990322 &7488 &PF &[4,110] &$-0.87_{+0.01}^{-0.02}$ &$1.27_{+0.13}^{-0.05}$ 
&102 &1.17 \\
990415 &7520 &V+T &[44,122] &$<-2.23$ &$<29.44$ &75 &1.54 \\
990518 &7575 &V+T &[177,291] &$-1.92_{+0.90}^{-0.92}$ &$154.57_{+15.76}^{-0.79}$ 
&110 &1.61 \\
991216 &7906 &V+T &[35,75] &$-2.35_{+0.03}^{-1.15}$ &$26.76_{+0.00}^{-6.25}$ 
&37 &1.23 \\
991229 &7925 &FV &[16,154] &$<-2.09$ &$<-24.29$ &132 &1.16 \\
000103 &7932 &V+T &[53,250] &$-1.90_{+0.90}^{-1.10}$ &$18.46_{+24.58}^{-7.22}$ 
&191 &1.39 \\
\enddata
\tablenotetext{(1)}{Abbreviations for time profile descriptions: PF = pure 
FRED (i.e., single smooth pulse), FV = FRED with multiple pulses near the 
peak, V+T = variability + tail.}
\tablenotetext{(2)}{First emission episode of trigger 5478.}
\tablenotetext{(3)}{Second emission episode of trigger 5478.}
\tablenotetext{(4)}{The longest duration GRB observed by BATSE ($\sim 3000$ 
seconds).}
\end{deluxetable}

The model was fit to the data using a Levenberg-Marquardt nonlinear 
least squares $\chi^{2}$ minimization algorithm. The algorithm was
modified to incorporate model variances rather than data variances 
in the computation of the $\chi^{2}$ statistic to avoid overweighting 
data points with strong downward Poisson fluctuations (\cite{ford95}). 
We performed a set of Monte Carlo simulations to test the accuracy of 
our fitting method. We found a bias in the distribution of fitted 
slopes that was hinged on the correlation of the ($\beta$,$t_{0}$) 
model parameters. The bias results when the value of $\tau_{1}$ is 
too far out in the tail of the power-law. In this situation, the 
curvature of the power-law decay is undersampled and results in a 
broad $\chi^{2}$ minimum. The broad $\chi^{2}$ minimum is most easily 
illustrated by plotting the joint confidence intervals between $\beta$ 
and $t_{0}$. For example, Figure 1 shows the $\Delta \chi^{2}$ contours 
for the fit to GRB970925 with $\Delta \chi^{2} =$ 2.3, 6.2, and 11.8, 
corresponding to the $68\%$ ($1\sigma$), $95\%$ ($2\sigma$), and $99\%$ 
($3\sigma$) confidence levels for two parameters of interest 
(\cite{press92}). 

\begin{center}
   \psfig{figure=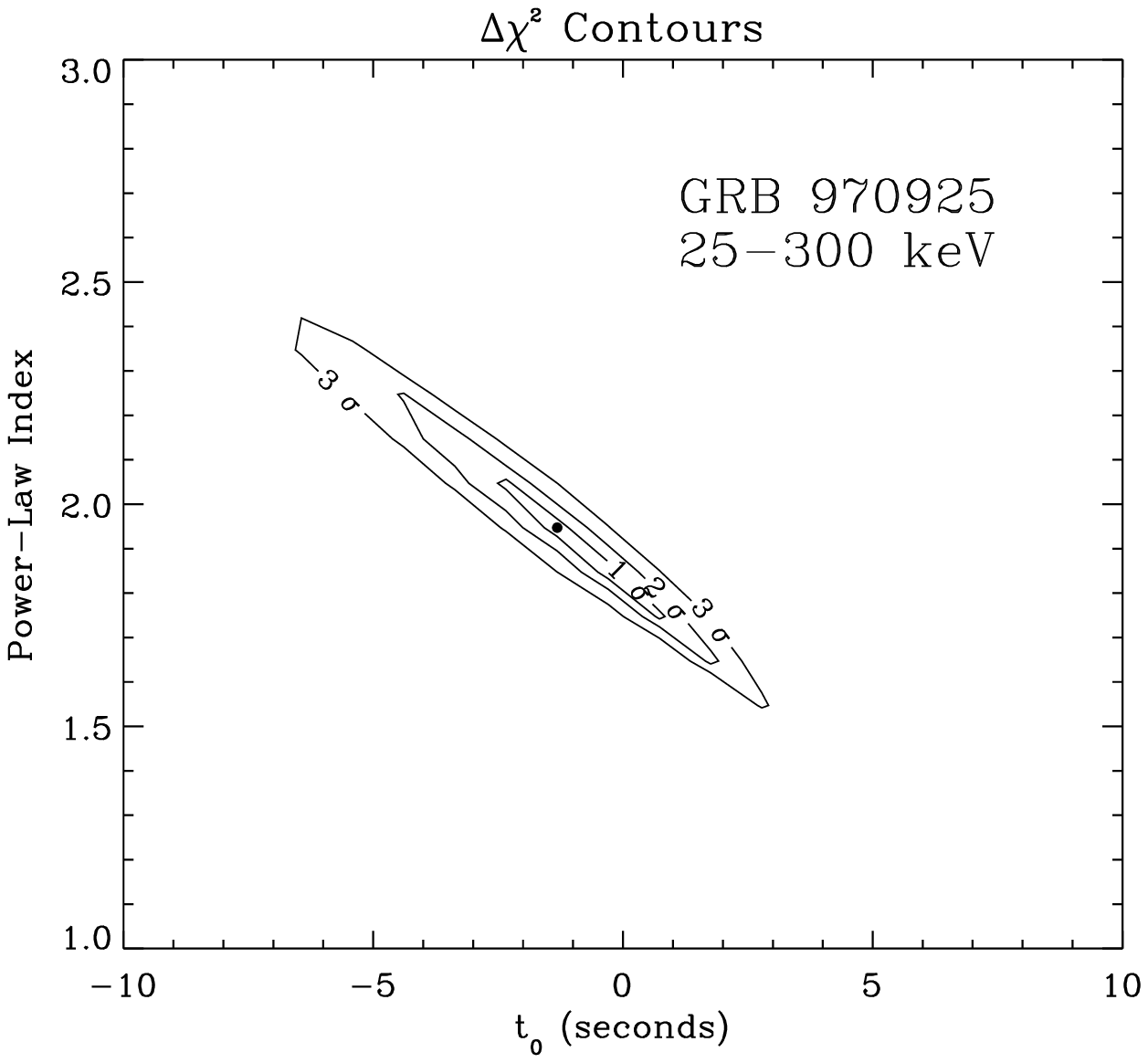,width=8cm}
   \figcaption[fig1.ps]{$\Delta \chi^2$ contour plot for GRB970925 that 
   shows the correlation between $\beta$ and $t_{0}$. Elliptical contours 
   are the $1\sigma$ ($68\%$), $2\sigma$ (95\%), and $3\sigma$ (99\%) joint 
   confidence intervals. Values of $\beta$ and $t_{0}$ corrresponding to the 
   fitted $\chi^{2}$ minimum are indicated by the filled circle. 
   \label{figure1}}
\end{center}

Examples of three burst decays from our sample are displayed in Figure 
2. The dashed lines indicate the best-fit power-law model for each event 
as listed in Table 1. The temporal fit parameters for all events in our 
sample are given in Table 1. The uncertainties in the $\beta$ and $t_{0}$ 
parameters quoted in Table 1 reflect the projection of the $68\%$ 
confidence contours onto the axis of the parameter of interest, and, 
in nearly all cases are larger than the uncertainties obtained from 
the covariance matrix in the Levenberg-Marquardt algorithm. It is 
important to point out that modeling of the temporal decay of afterglow 
measurements at very late times after the burst (e.g., days, weeks, and 
months) in, for example, the optical band does not suffer from this bias 
because the value of $t_{0}$ is typically set to the trigger time of the 
burst, i.e., a very good approximation to the true value of $t_{0}$ 
relative to the time of the fit interval (e.g., \cite{fruchter99}). 
In the case of the early decays in GRBs however, the fit is very 
sensitive since we are fitting so close in time to the burst trigger.

For completeness we also modeled the decay interval in each event 
with an exponential function of the form 
$R_{e}(t_{i}) \sim \exp[-(t_{i}-\tau_{1})/\tau_{e}]$ with the amplitude 
and exponential decay constant $\tau_{e}$ as free parameters. We find 
that only 12 of the 41 fits resulted in a lower reduced $\chi^{2}$ value, 
$\chi_{r}^{2}$, than that of the power-law model. The largest value of 
$\Delta \chi_{r}^{2}$ of these events was only 1.15 while most other 
events had $\Delta \chi_{r}^{2} \sim 0.3$ or less, indicating that the 
power-law is nearly as good a fit as the exponential. For events in 
which the exponential model was a poor fit, the power-law fits were 
strongly favored with $\Delta \chi_{r}^{2}$ values as high as 7.5. 
Our results are consistent with that of a similar study for a small 
number of GRBs performed early in the BATSE mission 
(\cite{schaefer95}).

\begin{center}
   \psfig{figure=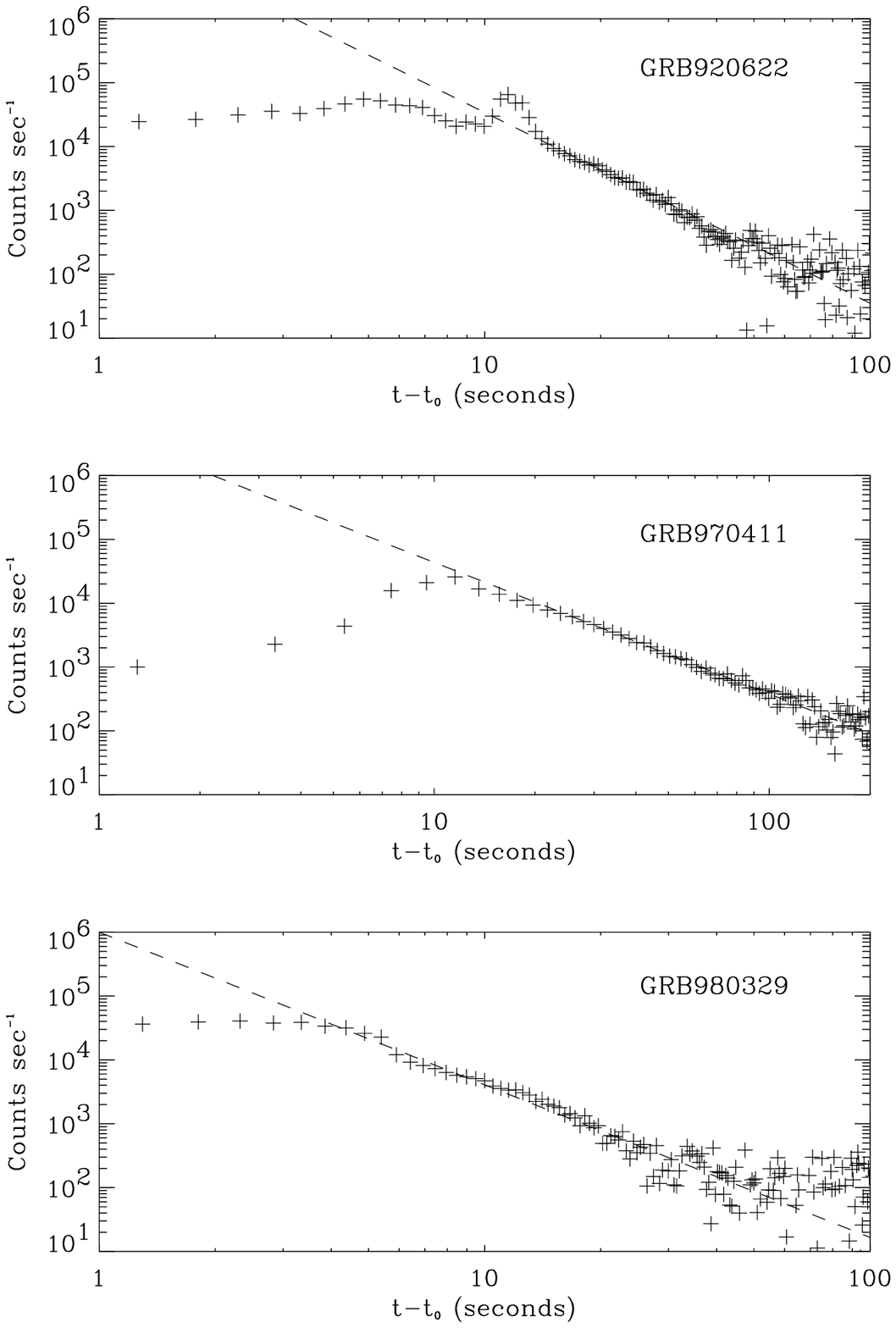,bbllx=119bp,bblly=108bp,bburx=504bp,%
   bbury=666bp,width=8cm}
   \figcaption[fig2.ps]{A logarithmic plot of the time histories of 
   three events in the 25-300 keV range. The dashed line is the best-fit 
   power-law model for each burst. The time intervals for each fit are 
   listed in the fourth column in Table 1. \label{figure2}}
\end{center}

\subsection{Spectral Modeling}

Nearly all GRB spectra are adequately modeled with a low and high-energy
power-law function smoothly joined over some energy range within the BATSE 
energy bandpass (\cite{band93}, \cite{preece00}). Curvature in the spectrum 
is almost always observed, although on rare occasions a broken power-law 
(BPL) model is a better represention of the data (\cite{preece98}). Spectra 
of X-ray afterglows in the 2-10 keV range observed with {\it BeppoSAX} are 
best fit with a single power-law, with spectral indices that range from 
$-1.5$ to $-2.3$ (\cite{costa00}). Recently it has been noted that the 
breaks in the synchrotron spectrum may not be sharp, but rather smooth
(\cite{granot01}). Therefore we chose two spectral forms to model the 
spectra of the gamma-ray tails: a single power-law as a baseline function, 
and a smoothly broken power-law (SBPL). The SBPL was chosen to enable direct 
comparison with the spectral form of the synchrotron shock model.

Because we are interested in the spectral behavior during late times 
of the burst, the CONT data-type from individual detectors is the 
optimum choice of available data-types from the LADs. CONT affords 
the best compromise between temporal and energy coverage, with 16 
energy channels and 2.048 s time resolution. Coarse temporal and 
energy bins are required as we are dealing with a signal that 
continuously decays with time. 

We model the photon spectrum (photons s$^{-1}$ cm$^{2}$ keV$^{-1}$) 
using the standard deconvolution and the Levenberg-Marquardt nonlinear 
least-squares fitting algorithm that incorporates model variances. The 
spectra were modeled using CONT channels 2-14, which covered the energy 
range of $\sim 30$-1800 keV. Count spectra from the two brightest detectors 
(i.e., the two detectors with the smallest source angles to the LAD normal 
vector) were generally used to make the fits. In some cases, the source 
angles differed substantially ($>20^{\circ}$) resulting in a normalization 
offset in the count spectra between the two detectors. The data from the 
two detectors were fit jointly with a multiplicative effective area 
correction term in the spectral model. For bursts in which the effective 
area correction was small ($\lax 5\%$), we summed the CONT count rates 
from the individual detectors to maximize the count statistics. These 
were cases in which the source angles of the two detectors differed by 
only a few degrees. Detectors with angles to the source exceeding 
$60^{\circ}$ or with strong signal from sources such as Vela X-1 or 
Cyg X-1 were excluded from the fit.

\begin{deluxetable}{lccccccc}
   \footnotesize
   \tablecaption{Time-Integrated Spectral Fits: Smoothly Broken 
      Power-Law \tablenotemark{(1)} \label{tbl-2}}
   \tablehead{
   \colhead{GRB} & \colhead{$\alpha$\tablenotemark{(2)}} & 
   \colhead{$\alpha^{\prime}$\tablenotemark{(2)}} & 
   \colhead{$E_{b}$ (keV)} & \colhead{$\chi^{2}/d.o.f.$} & 
   \colhead{$\Delta$} & \colhead{$p$} & \colhead{category\tablenotemark{(4)}}
   }
   \startdata
   910927 &$0.06\pm0.04$ &$-3.27\pm0.17$ &$158\pm7$ &1.84 
	&$3.33\pm0.17$ &$6.54\pm0.17$ &\nodata \\
   920218 &$-0.69\pm0.01$ &$-1.45\pm0.05$ &$175\pm7$ &2.37 
	&$0.76\pm0.15$ &$2.90\pm0.05$ &\nodata \\
   920502 &$-0.10\pm0.05$ &$-1.96\pm0.20$ &$183\pm20$ &1.22
	&$1.86\pm0.21$ &$3.92\pm0.20$ &\nodata \\
   920622 &$-0.49\pm0.04$ &$-1.48\pm0.26$ &$263\pm68$ &1.97 
	&$0.99\pm0.26$ &$2.96\pm0.26$ &(i),(iii) \\
   920801 &$-0.23\pm0.05$ &$-0.98\pm0.15$ &$252\pm70$ &1.13 
	&$0.75\pm0.16$ &$1.96\pm0.15$ &(iii) \\
   930612 &$-0.01\pm0.17$ &$-1.46\pm0.12$ &$108\pm19$ &0.49
	&$1.45\pm0.21$ &$2.92\pm0.12$ &\nodata \\
   931223 &$-0.24\pm0.06$ &$-1.39\pm0.13$ &$141\pm12$ &1.28 
	&$1.15\pm0.14$ &$2.79\pm0.13$ &\nodata \\
   940419b &$-0.52\pm0.17$ &$-1.66\pm0.55$ &$140\pm73$ &1.15
	&$1.14\pm0.58$ &$3.32\pm0.55$ &(i) \\
   941026 &$-0.24\pm0.03$ &$-1.55\pm0.08$ &$150\pm6$ &2.32
	&$1.32\pm0.09$ &$3.10\pm0.08$ &\nodata \\
   960530(1) &$-0.10\pm0.18$ &$-1.48\pm0.22$ &$128\pm33$ &1.59 
	&$1.47\pm0.28$ &$2.96\pm0.22$ &\nodata \\
   960530(2) &$-0.56\pm0.04$ &$-1.88\pm0.37$ &$203\pm25$ &1.54
	&$1.33\pm0.37$ &$3.76\pm0.37$ &(i)\tablenotemark{(5)} \\
   970411 &$-0.35\pm0.04$ &$-1.10\pm0.08$ &$228\pm41$ &1.60 
	&$0.75\pm0.09$ &$2.20\pm0.08$ &(iii) \\ 
   970925 &$-0.02\pm0.32$ &$-1.28\pm0.13$ &$101\pm32$ &1.10
	&$1.26\pm0.35$ &$2.56\pm0.13$ &(iii) \\
   971208 &$-0.55\pm0.02$ &$-2.03\pm0.04$ &$179\pm6$ &2.02
	&$1.48\pm0.04$ &$4.06\pm0.04$ &\nodata \\
   980301 &$-0.55\pm0.22$ &$-1.24\pm0.13$ &$76\pm34$ &0.80
	&$0.69\pm0.13$ &$2.48\pm0.13$ &(i),(iii) \\
   981203 &$0.29\pm0.05$ &$-0.59\pm0.01$ &$124\pm8$ &2.28 
	&$0.88\pm0.05$ &\nodata\tablenotemark{(3)} &(iii) \\
   990102 &$0.53\pm0.16$ &$-1.82\pm0.15$ &$121\pm14$ &1.03
	&$2.35\pm0.22$ &$3.64\pm0.15$ &\nodata \\
   990220 &$0.64\pm0.24$ &$-1.36\pm0.08$ &$94\pm13$ &5.79
	&$2.00\pm0.25$ &$2.72\pm0.08$ &\nodata \\
   990316 &$-0.58\pm0.04$ &$-1.52\pm0.09$ &$145\pm18$ &3.44
	&$0.94\pm0.10$ &$3.04\pm0.09$ &\nodata \\
   990518 &$-0.52\pm0.04$ &$-1.45\pm0.20$ &$174\pm22$ &2.07
        &$0.93\pm0.20$ &$2.90\pm0.20$ &(i) \\
   \enddata
   \tablenotetext{(1)}{For most events the fluence interval is 
      the same interval used in making the temporal fit.}
   \tablenotetext{(2)}{Here, $\alpha$ and $\alpha^{\prime}$ are the
      indices of the of the spectral energy flux, $F_{\nu}$
      (erg s$^{-1}$ cm$^{-2}$ keV$^{-1}$), i.e. $\alpha=\alpha_{\rm low}+1$ 
      and $\alpha^{\prime}=\alpha_{\rm high}+1$.} 
   \tablenotetext{(3)}{Note that the value of $\alpha$ is consistent with 
      the spectral slope below $\nu_{c}$ in the fast-cooling mode and the
      spectral slope below $\nu_{m}$ in the slow-cooling mode. In the former
      case, $p$ is undetermined. For slow-cooling, $p=2.18$.}
   \tablenotetext{(4)}{Characterisitc signatures of the synchrotron spectrum
      as described in $\S 3.3$ of the text.}
   \tablenotetext{(5)}{Within 1.5 sigma}
   \tablecomments{Uncertainties in the model parameters are taken from 
      the covariance matrix.}
\end{deluxetable}

\begin{deluxetable}{lcccc}
   \footnotesize
   \tableheadfrac{0.20}
   \tablecaption{Time-Integrated Spectral Fits: Single 
      Power-Law\tablenotemark{(1)} \label{tbl-3}}
   \tablewidth{420pt}
   \tablehead{
   \colhead{GRB} & \colhead{Trigger} & 
   \colhead{$\alpha_{p}$\tablenotemark{(2)}} & 
   \colhead{$\chi^{2}/d.o.f.$} & \colhead{$p$\tablenotemark{(3)}}
   }
   \startdata
   910602 &257  &$-0.52\pm0.01$ &1.05 &$1.04\pm0.01$ \\
   911016 &907  &$-1.77\pm0.22$ &1.22 &$3.54\pm0.22$ \\
   920813 &1807 &$-1.01\pm0.02$ &13.2 &$2.02\pm0.02$ \\
   930131 &2151 &$-0.89\pm0.11$ &0.89 &$1.78\pm0.11$ \\
   951104 &3893 &$-1.47\pm0.05$ &4.96 &$2.94\pm0.05$ \\
   970302 &6111 &$-0.73\pm0.10$ &1.13 &$1.46\pm0.10$ \\ 
   990322 &7488 &$-0.42\pm0.06$ &1.51 &$0.84\pm0.06$ \\
   990415 &7520 &$-1.28\pm0.16$ &1.19 &$2.56\pm0.16$ \\
   000103 &7932 &$-1.99\pm0.43$ &1.34 &$3.98\pm0.43$ \\
   \enddata
   \tablenotetext{(1)}{For most events the fluence interval is the 
      same interval used in making the temporal fit.}
   \tablenotetext{(2)}{Power-law index is the index of the spectral 
      energy flux, $F_{\nu}$ (erg s$^{-1}$ cm$^{-2}$ keV$^{-1}$).}
   \tablenotetext{(3)}{Here, $p = -2\alpha_{p}$.}
   \tablecomments{Uncertainties in the model parameters are taken from 
	the covariance matrix.}
\end{deluxetable}

The free parameters of the power-law spectral model are the amplitude 
and the power-law index $\alpha_{p}$. The free parameters of the BPL 
and SBPL are the amplitude, low-energy index $\alpha_{\rm low}$, 
high-energy index $\alpha_{\rm high}$, and the break energy $E_{b}$. 
The slopes of the spectral energy flux, $F_{\nu}$ are readily obtained 
from the simple relations: $\alpha = \alpha_{\rm low} + 1$ and 
$\alpha^{\prime} = \alpha_{\rm high} + 1$. 

For the decay emission of each burst, we modeled the time-integrated
spectrum defined over a time interval that was either the same as or
shorter in length than the time interval used in making the temporal 
fits. In all cases, the time interval was restricted {\it to the region 
of the burst during the power-law decay}. Shorter time intervals were 
used for events with weaker signal-to-noise. For the PF class bursts, 
however, we selected the entire burst emission (spectra of the burst 
intervals starting at the peak gave nearly identical parameter values). 
For the majority of events, the SBPL model was preferred over the single 
power-law model. Summarized in Table 2 are the spectral fit parameters 
for events where the SBPL was the better choice of model based on the 
$\chi^{2}$ statistic. In 6 bursts, however, the fitted value of the 
high-energy slope was unusually steep ($\lax -4.0$). Preece et al.\ (1998) 
pointed out that spectral models with curvature can sometimes overestimate 
the steepness of the high-energy slope, depending on how well the data 
tolerate curvature (see figure 1 in \cite{preece98}). The broken power-law 
model was therefore used for these 6 events, resulting in slightly better 
reduced $\chi^{2}$ values and better constrained values of the fitted 
high-energy slope. This resulted in spectral parameters for a total of 
20 bursts. The reduced $\chi^{2}$ values are reasonable, although a few 
bursts for which joint fits were made tended to give slightly larger values 
($\chi^{2}/d.o.f. \gax 2$). Given in the table are the fitted values of the 
spectral indices, the break energy, and their uncertainties from the covariance
matrix. Also given is the difference in spectral slope across the break energy,
 $\Delta = \mid\alpha^{\prime} - \alpha\mid$, and the value of $p$ 
calculated from the high-energy spectral slope.

Of the remaining bursts, 9 events were best represented by the single 
power-law function. The best fit parameters and the corresponding value 
of $p$ are presented in Table 3. The spectral fits for the remaining 12 
events resulted in poor $\chi^{2}$ values and poorly constrained parameter 
values, regardless of the choice of spectral model. These are clearly cases 
when the counting statistics are too poor to constrain the model parameters
and therefore were excluded. The results in Table 3 should be interpreted 
with some degree of caution. These events may be cases in which the flux 
level was to low, causing the break in the spectrum to be washed out in 
the counting noise. In such cases, the single power-law will often be 
adequate to model the spectrum, even though the true burst spectrum may 
contain a break.

A careful inspection of Table 2 immediately allows us to identify 
high-energy afterglow candidates based on three characteristic signatures 
of the synchrotron spectrum which we categorize as the following: (i) in 
the fast-cooling mode, the spectral slope below the high-energy break 
($\nu_{\rm m}$) is {\it always} $-1/2$ for radiative or adiabatic 
evolution, as seen from equations 3 and 4, (ii) in the slow-cooling mode, 
the change in spectral slope across the high-energy break ($\nu_{\rm c}$) 
is {\it always} $1/2$, as seen from equation 5, and (iii) the electron 
energy index, $p$, calculated from the measured spectral slope should
have a value in the range $2.0 \le p \le 2.5$, the typical range derived 
from afterglows observed at X-ray, optical, and radio wavelengths. 

Applying these criteria, we label events with these properties in the 
last column of Table 2. We thus immediately identify several fast-cooling 
candidates: GRB920622, GRB940419b, GRB960530(2), GRB980301. Each of these 
events has a value of $\alpha$ within one-sigma of $-0.5$ and a value 
of $p$ similar to those found for afterglows. GRB970411, GRB971208,
GRB990316, and GRB990518 are only marginally consistent with
fast-cooling, having larger $p$ values and reduced $\chi^2$'s.
None of the events in Table 2 are consistent (within one-sigma) 
with $\Delta = 0.5$, suggesting no slow-cooling candidates (however,
5 events [GRB920218, GRB920622, GRB920801, GRB940419b, and GRB980301]
have values within two-sigma). A total of 9 bursts in Table 2 are 
ruled out as high-energy afterglow candidates because their spectral 
parameters bear no resemblance to the fast or slow-cooling synchrotron 
spectrum. One event of notable interest is GRB981203, which has $\alpha$ 
and $\alpha^{\prime}$ values consistent with a cooling break $\nu_{\rm c}$, 
as opposed to $\nu_{\rm m}$, in the fast-cooling spectrum. This implies a 
$\nu_{\rm m}$ break above $\sim 2$ MeV, while the value of $p$ remains 
unconstrained by the data. In $\S 4$ the early high-energy afterglow 
candidates are discussed in greater detail.

Obviously, for the single power-law events listed in Table 3 we 
have less spectral information. The value of $p$ given in Table 
3 is derived from $\alpha_{p}$ under the assumption that $\alpha_{p}$ 
is the slope above the break, for fast or slow-cooling. Clearly this 
need not be the case. A case in point is GRB910602, which has 
$\alpha_{p} = -0.52 \pm 0.01$, a value consistent with the spectral 
slope of fast-cooling for $\nu_{\rm c} < \nu < \nu_{\rm m}$. In this 
interpretation the cooling break, $\nu_{\rm c}$, would be below the 
BATSE window and $\nu_{\rm m}$ above. The value of $p$ would be 
undetermined from the data. Scanning the values of $p$ given in 
Table 3, we find $p = 2.56 \pm 0.16$ for GRB990415, a typical 
value for afterglows. This suggests that the measured slope 
$\alpha_{p} = -1.28 \pm 0.16$ could be the slow or fast-cooling 
high-energy slope. What values of the spectral slope do we expect 
to observe below $\nu_{\rm c}$ for slow-cooling? If we assume a value 
of $p = 2.5$, then the calculated slope below $\nu_{\rm c}$ is 
$\alpha = -(p-1)/2 = -0.75$. We find one burst, GRB970302, with 
$\alpha_{p} = -0.73 \pm 0.10$, consistent with the expected value 
if $p = 2.5$. 

\begin{center}
   \psfig{figure=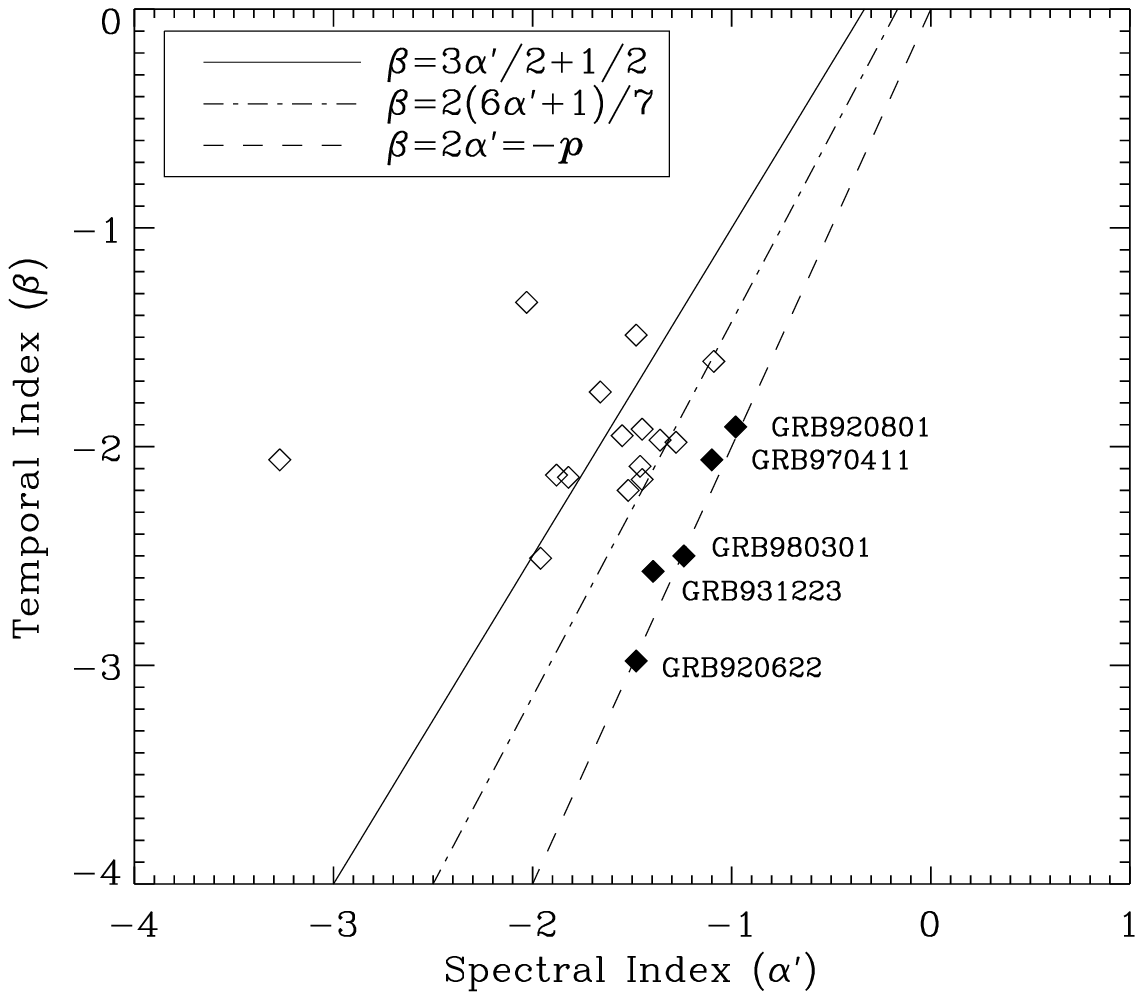,width=3.5truein}
   \figcaption[fig3.ps]{A plot of high-energy spectral index vs.\ 
   temporal index for the twenty events listed in Table 2. Also 
   plotted are the linear relationships expected from the evolution 
   of an adiabatic spherical blast wave ({\it solid line, dash-dot line}) 
   and a jet ({\it dashed line}). The five bursts labeled on the plot 
   ({\it fill diamonds}) are within one-sigma of the 
   $\beta = 2\alpha^{\prime}$ line. \label{figure3}}
\end{center}

An additional constraint we can apply to the data is a comparison of 
the measured temporal slopes with their expected values derived from 
the measured spectral indices given in the expressions in equation 6. 
A plot of temporal vs.\ spectral index for the data in Table 2 is shown 
in Figure 3. Here, the spectral index is the high-energy spectral energy 
index, $\alpha^{\prime}$, in the third column of Table 2. For comparison 
with the models, we plot the possible linear relationships between 
$\alpha^{\prime}$ and $\beta$ given in equation 6. Note that this plot 
should be interpreted with a certain degree of caution. The expected 
values of $\alpha^{\prime}$ are somewhat restricted by the possible 
range of $p$ values between 2.0 and 2.5 predicted by Fermi acceleration 
models (e.g., \cite{gallant99}, \cite{gallant00}). Interestingly, however, 
five events ({\it filled diamonds}) are consistent with the $\beta = -p$ 
line for adiabatic jet evolution. We address this implication in detail 
in $\S 4$ and $\S 5$.

A similar plot is shown in Figure 4 for the single power-law fits from 
Table 3. In general, all but one event appear consistent with the relations 
between the temporal and spectral indices expected from external shocks. 
Thus, closer inspection of the spectral parameters (e.g., $\alpha_{p}$ and 
$p$) in Table 3 is required to establish if these are viable high-energy 
afterglow candidates (see $\S 4$).

\subsection{Spectral Evolution: Color-Color Diagrams}

The evolution of the synchrotron spectrum is unique for a given 
hydrodynamical evolution of the blast wave. This evolution can be
traced in a graphical form using a Color-Color Diagram (CCD). The 
CCD method is a model-independent technique that characterizes the 
spectral evolution of the burst over a specified energy range. With 
this method, a comparison of spectral evolution patterns among GRBs 
can be made in addition to a comparison with patterns expected from 
the evolution of the synchrotron spectrum. 

\begin{center}
   \psfig{figure=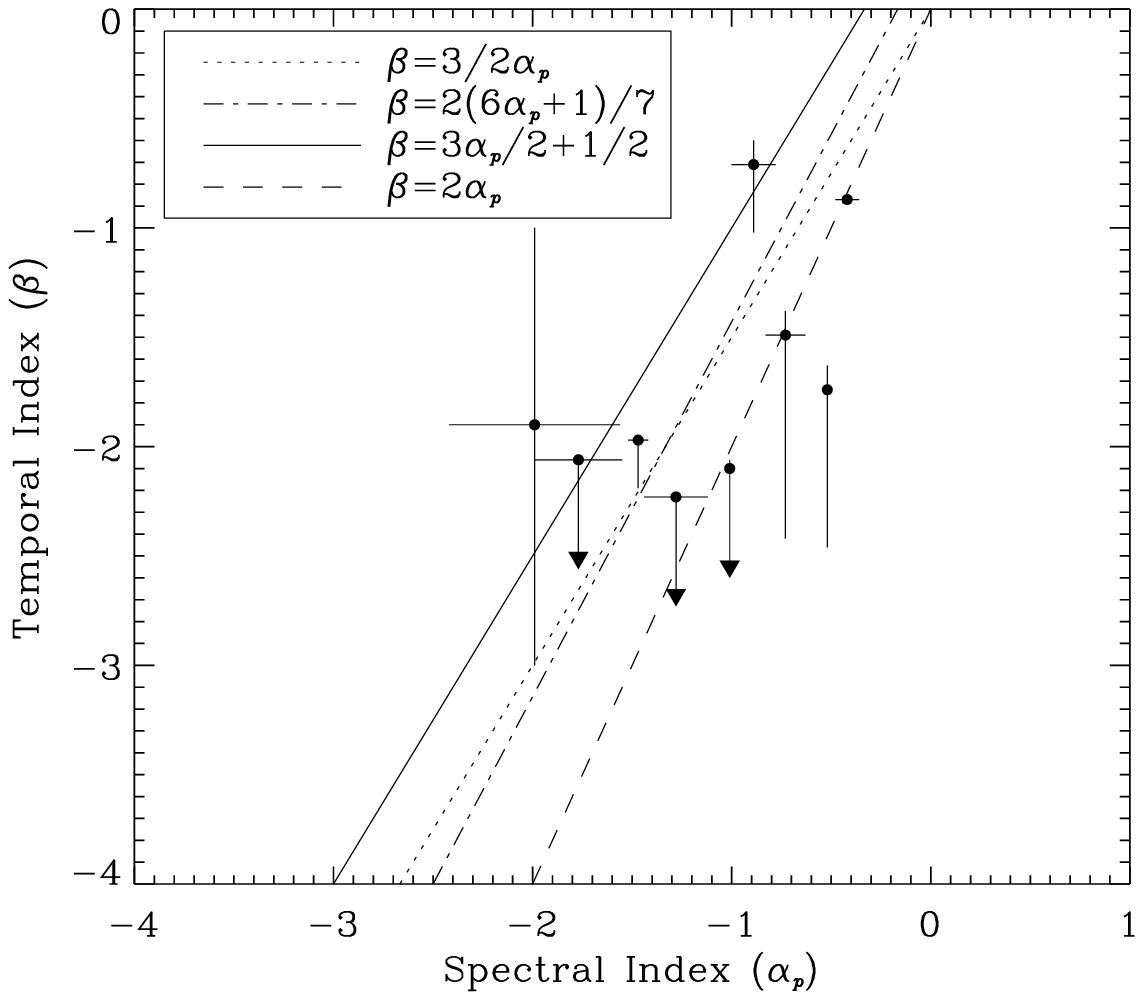,width=3.5truein}
   \figcaption[fig4.ps]{A plot of spectral index vs. temporal index for 
   the 9 events listed in Table 3. The asymmetric uncertainties in the 
   temporal indices reflect the $68\%$ joint confindence intervals between 
   $\beta$ and $t_{0}$. Thus the uncertainties for bursts with highly 
   elongated $\Delta\chi^{2}$ contours are indicated by downward arrows. 
   Lines indicate the linear relationships expected in the synchrotron 
   afterglow spectrum: $\nu_{m} < \nu < \nu_{c}$ in slow cooling, 
   $\beta = 3 \alpha_{p} / 2$ ({\it dotted line}), $\nu_{m} < \nu$ in 
   adiabatic fast-cooling or $\nu_{c} < \nu$ in slow cooling, 
   $\beta = 3 \alpha_{p} / 2 + 1/2$ ({\it solid line}), and for 
   $\nu_{m} < \nu$ in radiative fast cooling, 
   $\beta = 2 (6\alpha_{p} + 1)/7$ ({\it dash-dot}). Also plotted is 
   the $\beta = 2\alpha_{p}$ line for adiabatic jet evolution. Since 
   we do not observe a break in the spectrum for these events and thus 
   can not distinguish if the observed power-law index is the low or 
   high-energy index, we plot all possible relationships between the 
   spectral and temporal index. \label{figure4}}
\end{center}

The CCD is a plot of the hard color vs.\ the soft color, where the 
hard and soft colors are defined as the hardness ratios (i.e., 
ratios of the count rates) between (100-300 keV/50-100 keV) and 
(50-100 keV/25-50 keV), respectively. To construct the CCDs we use 
the count rates in the three lowest (25-300 keV) of the four broad 
energy channels from the DISCSC data. We select a time interval large 
enough to cover most of the burst emission until the statistical noise 
begins to dominate. These bins are identified by hardness ratios with 
two-sigma upper limits. We also fold the fast and slow-cooling broken 
power-law synchrotron spectra of a spherical blast wave through the LAD 
detector response to obtain the expected count spectrum (shown as the 
dashed and solid lines, respectively, in Figures 5-8). We assume the 
fast-cooling spectrum is radiative, with $p=2.4$, $\alpha_{\rm low} = -1.5$, 
and allow $E_{\rm m}$ to evolve from $220 \rightarrow 25$ keV. For the 
slow-cooling spectrum, we also assume $p=2.4$ and the same evolution 
for $E_{\rm c}$. 

\begin{figure*}
\centerline{\psfig{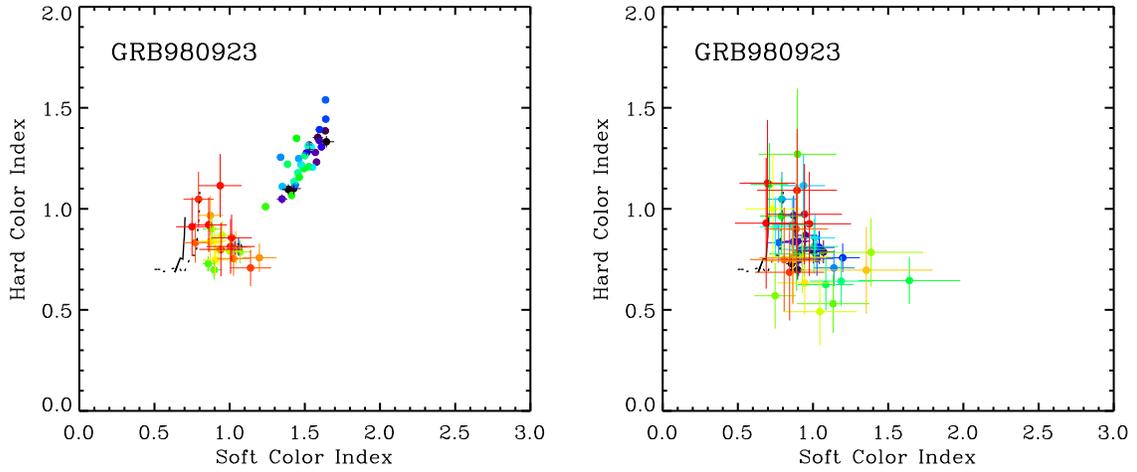}}
   \figcaption[fig5.cps]{{\it Left panel:} Color-color diagram of 
   GRB980923 variability episode {\it and} tail emission (0-70 seconds). 
   The crescent-like pattern is the variable emission while the decoupled 
   cluster of points is the tail emission. {\it Right panel:} Color-color 
   diagram of the tail emission only, covering a much longer time interval 
   (40-100 seconds). Also plotted are the evolution patterns expected from 
   the slow ({\it solid}) and fast-cooling ({\it dashed}) synchrotron 
   spectrum. \label{figure5}}
\end{figure*}

Giblin et al.\ (1999a) have shown that the tail emission from 
GRB980923 resembles that of afterglow synchrotron emission due 
to an external shock. To illustrate the usefulness of the CCD 
technique, we show in Figure 5 the CCD for GRB980923. In this 
representation, the time evolution of the burst is preserved by 
a color sequence of the hardness ratios, with black/violet/blue 
signaling the onset of the burst and yellow/red signaling the end 
of the burst. The left panel shows the CCD for the time interval 
that brackets the entire burst (variability + tail). The variable 
emission of the burst shows a crescent-like pattern decoupled 
from a cluster of points that represent the tail of the burst. 

The crescent pattern is typical among GRBs (\cite{kouveliotou93b},
\cite{giblin99b}), however the clustering is less common. The 
crescent track exhibits a sawtoothing of soft-hard-soft evolution, 
indicative of the spectral behavior of the individual pulses that 
comprise the main burst emission. The pattern drastically changes 
when the variability ceases and the tail becomes visible. The tail 
cluster overlaps the region of the two-color plane that contains 
the evolution of the slow and fast-cooling synchrotron spectrum. 
This is best illustrated in the right panel of Figure 5, where 
the CCD is constructed from a longer time interval in the tail
only. Unfortunately the CCD pattern of the tail is not completely 
resolved due to the increasingly large uncertainties in the 
hardness ratios that arise from the decreasing flux level. 
However, the points do lie in the correct region of the diagram. 
This decoupling of the points in the model-independent CCD is 
clear evidence for two distinct spectral components observed 
in a GRB. 

Figure 6 shows the CCDs for the four fast-cooling candidates 
identified based on their spectral parameters in Table 2. The 
pattern for GRB920622 bears a striking resemblance to that of 
GRB980923 in the left panel of Figure 5. Like GRB980923, this 
burst contains a period of variability followed by a very smooth 
emission tail. The crescent pattern that arises from the variable 
part of the burst is clearly visible and spans nearly the same 
range of soft and hard color indices. However, the clear 
discontinuity between the burst emission and the tail emission 
in the CCD of GRB980923 is not as pronounced in the CCD of GRB920622. 
Nonetheless, the tail emission (denoted by the orange and red points) 
lies in the same region as those of GRB980923 and the synchrotron 
afterglow spectrum. The CCD pattern for GRB949419b (in the PF class) 
is similar, but appears more cluster-like in the region of synchrotron 
evolution. 
\begin{figure*}[tbp]
\centerline{\psfig{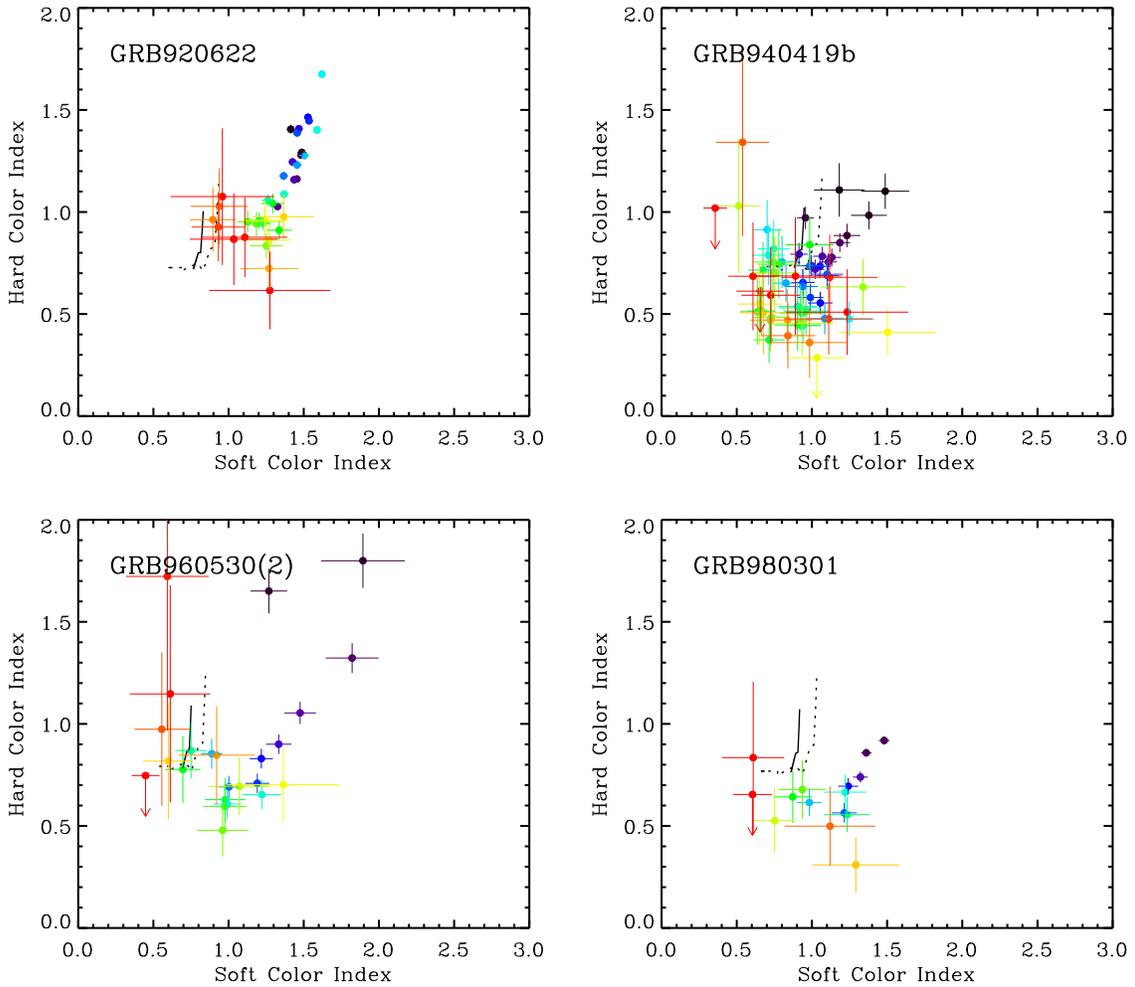}}
   \figcaption[fig6.cps]{Color-color diagrams for the four fast-cooling
   candidates from Table 3. Arrows indicate two-sigma upper limits. 
   \label{figure6}}
\end{figure*}

\begin{figure*}
\centerline{\psfig{figure=fig7.cps,width=4.8truein,bbllx=108bp,bblly=270bp,%
   bburx=522bp,bbury=655bp}}
   \figcaption[fig7.cps]{Color-color diagrams for the marginal fast-cooling 
   candidates from Table 3. Arrows indicate two-sigma upper limits.
   \label{figure7}}
\end{figure*}

Another burst of interest in the PF category is GRB960530. This event 
consists of two FREDs separated by $\sim 200$ seconds. The 2nd FRED only 
has about half of the peak intensity as the first. The CCD for the 2nd 
FRED is seen in Figure 6. The episode begins very hard on the rise and
evolves through the synchrotron spectrum during the decay. Interestingly, 
although the 1st episode of GRB960530 is also a FRED, its color-color 
diagram (Figure 8) shows a broad crescent pattern that evolves much 
farther away from the synchrotron pattern. This may be a case where 
the external shock is clearly decoupled in time from the GRB.

The last event in Figure 6, GRB980301, shows an intriguing pattern that
closely resembles the evolution pattern of the synchrotron spectrum. The 
evolution is mainly shaped like a reverse ``L'', but marginally offset to 
higher soft color values and lower hard color values. In this case it is 
difficult to argue in favor or against the synchrotron model.

Figure 7 shows the CCDs of the four events from Table 3 that are only
marginally consistent with fast-cooling. GRB970411 and GRB990518 show
similar patterns that resemble those in Figure 6, but the consistency
with the synchrotron pattern is weak. GRB971208 shows little evolution
and a cluster of points partially overlapping the synchrotron region. 
GRB990316 shows a nearly identical pattern to that of GRB980301.

A total of 23 events from our sample showed CCDs inconsistent with 
the evolution of the synchrotron spectrum. Rather, most of these events 
showed the crescent pattern that are common among GRBs, as depicted in 
Figure 8. These events span a much larger range of hard and soft colors 
than expected from the synchrotron emission alone. Others were too weak 
to distinguish a pattern.

\section{High-Energy Afterglow Candidates}
We identify a total of 8 from our sample of 40 events as high-energy
afterglow candidates based on their observed spectral parameters and 
color-color diagrams. Each burst is discussed in detail below.

\subsubsection{GRB\ 910602}

The observed spectral slope for GRB910602 is consistent with the spectral 
slope below $\nu_{\rm m}$ in the fast-cooling spectrum, although the slope 
below $\nu_{\rm c}$ in slow-cooling can not be ruled out, implying a value 
of $p = 2$. A series of time-resolved fits with a uniform time resolution 
of 4.096 s revealed no softening of the spectrum, i.e, the slope remained 
constant with $\alpha_{p} \sim -0.5$ throughout the tail. Applying the 
relations in equation 6, for slow-cooling we expect $\beta = -0.78 \pm 0.01$ 
and, from equations 4 and 5, for fast-cooling we expect $\beta = -4/7$ 
(radiative) or $\beta = -1/4$ (adiabatic). These values do not agree with 
the measured value $\beta = -1.74_{+0.11}^{-0.72}$. Although the spectrum 
appears to be consistent with that of the synchrotron spectrum, the 
evolution does not appear to be consistent with the evolution of a 
spherical blast wave.

\subsubsection{GRB\ 920622}

The time history of this burst bears a striking resemblance to that 
of GRB980923 reported by Giblin et al.\ (1999). Initially the burst 
is highly variable, then at $\sim 18$ s after the trigger time the 
burst enters a phase of smooth decay that lasts until $\sim 50$ s 
after the trigger. From Table 2, the time-integrated spectral fit 
suggests fast-cooling, with low-energy index $\alpha = -0.49 \pm 0.04$. 
From the high-energy index, a value of $p = 2.96 \pm 0.26$ is inferred. 
The value of $\Delta = 0.99 \pm 0.26$ is marginally consistent (within
two-sigma) with the expected value of 0.5 for slow-cooling. The 
time-integrated spectrum of the variable emission of the burst is in 
contrast with the fluence spectrum of the tail. The variable emission 
gives $\alpha_{\rm v} = -0.07 \pm 0.01$, 
$\alpha_{\rm v}^{\prime} = -1.50 \pm 0.04$, and $E_{{\rm v},b} = 370 \pm 12$ 
keV, suggestive of a spectral change near $\sim 18$ seconds. Note that the 
spectral parameters of the variable emission are {\it not} consistent 
with the synchrotron spectrum. We binned the tail emission into three 
time bins with S/N $\ge 45$ to model the spectral evolution, however 
the parameters were poorly constrained due to the steep nature of the 
flux decay. The spectral evolution of the variability + tail emission, 
however, can be seen in Figure 6. The tail of the burst appears 
consistent with the region of the diagram defined by the evolution of
the synchrotron spectrum. The measured temporal index of the tail is 
$\beta \le -2.98$, clearly inconsistent with the expected values for 
$\beta$ for $\nu < \nu_{\rm m}$. Interestingly, however, $\beta$ is 
nearly identical to the value of $p$ inferred from the high-energy 
slope, as expected for jet evolution.

\begin{figure*}
\centerline{\psfig{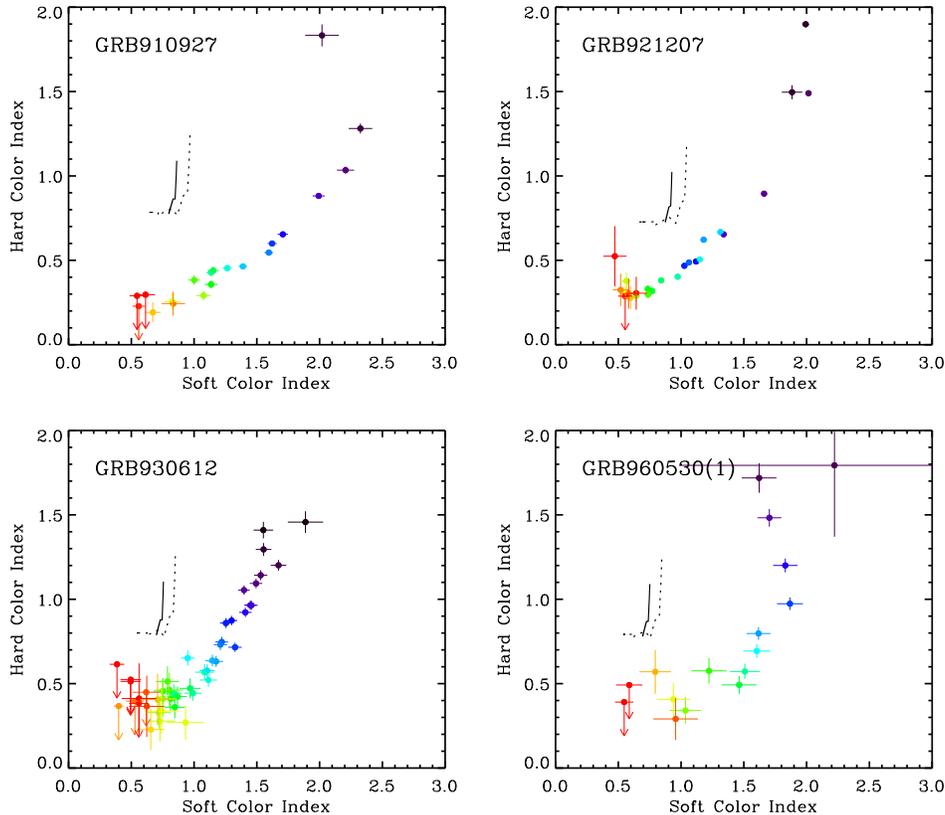}}
   \figcaption[fig8.cps]{A sample of color-color diagrams of GRBs that 
   are not consistent with the evolution of the synchrotron spectrum. 
   Arrows indicate two-sigma upper limits. \label{figure8}}
\end{figure*}

\subsubsection{GRB\ 940419b}

The smooth rise and decay structure of this burst place it in the PF
category. Like the tail emission of GRB920622, this burst also shows
a low-energy slope consistent with the fast-cooling spectral slope 
below $\nu_{\rm m}$. The measured temporal slope is $\beta \le -1.75$. 
While this slope is not consistent with the temporal index below
$\nu_{\rm m}$, it is consistent with the expected values of
$\beta = -2.56 \pm 0.55$ (radiative) and $\beta = -1.95 \pm 0.55$
(adiabatic) for $\nu > \nu_{\rm m}$, given the large uncertainties. 
We further binned the data in the tail to S/N $\ge 45$ and constrained 
the evolution of the break energy by holding the low and high-energy 
spectral indices fixed to their values derived from the time-integrated 
fit. We find $E_{b} \sim (t-t_{0})^{-1.21 \pm 0.15}$ for $t_{0}$ fixed 
at $-25.0$ s ($\chi^2/d.o.f.=7.52/8$). Additional fits with other larger 
values of $t_{0}$ gave slightly shallower indices, as expected if one 
compares the behavior to that of $\beta$ and $t_{0}$. For an adiabatic 
fast-cooling spherical blast-wave we expect the break energy to decay 
as $-1.5$, within two-sigma of our measured value. As seen from the CCD 
in Figure 5, the spectral evolution of this burst is very close to that 
of the synchrotron spectrum, although the rise of the burst tends to be 
somewhat harder in the soft color index than expected from evolution of 
the synchrotron spectrum alone.
  
\subsubsection{GRB\ 960530}

GRB960530 is of particular interest because of its striking temporal 
behavior. The burst has two distinct episodes of emission, each having 
a FRED-like time profile. The second episode, much weaker with a peak 
intensity less than half of the first, occurs $\sim 200$ seconds after 
the first. As seen in Table 2, the low-energy slope of the second 
episode is consistent with the fast-cooling slope below $\nu_{\rm m}$, 
however the value of $p = 3.76 \pm 0.37$ is large mainly because the
high-energy slope is not well-constrained. The value of $p$ derived 
for the first episode is not unreasonable, however the low-energy 
slope is roughly three-sigma away from the expected value of $-0.5$. 
The decay index for the second episode is $\beta \le -2.13$, not 
inconsistent with the expected values $\beta = -4.57 \pm 1.57$ and 
$\beta = -3.75 \pm 1.57$ for radiative and adiabatic fast-cooling, 
respectively. The CCD of the second emission episode of this burst 
(Figure 6) indicates that during the decay the emission evolves 
into the synchrotron spectrum. 

\subsubsection{GRB\ 970411}

From Table 2, the low-energy index of this burst is nearly four-sigma 
from the value $-0.5$ expected in the fast-cooling regime. However, it 
does have $p = 2.2 \pm 0.08$, consistent with typical afterglow values 
and particle acceleration models of relativistic shocks (e.g., 
\cite{gallant00}). Additionally, the change in slope, $\Delta$ is less 
than three-sigma from the expected value of $-0.5$ for the cooling break 
in the slow-cooling regime. For slow-cooling, we expect the temporal 
slope to be $\beta = -0.53 \pm 0.04$ for $\nu < \nu_{\rm c}$ and 
$\beta = -1.15 \pm 0.08$ for $\nu > \nu_{\rm c}$. For $\nu > \nu_{\rm m}$ 
in fast-cooling we expect $\beta = -1.60 \pm 0.08$ (radiative) and 
$\beta = -1.15 \pm 0.08$ (adiabatic). Our measured value of the decay, 
$\beta = -2.06_{+0.20}^{-0.21}$, is marginally consistent (within
two-sigma) with the radiative fast-cooling slope. More notably, it 
is consistent (within one-sigma) with the value of $p$. A series of 
spectral fits during the tail of the burst holding the low and 
high-energy spectral indices constant show that $E_{b}$ decays 
with time described by a power-law of the form 
$E_{b} \sim (t-t_{0})^{-0.96\pm0.26}$ for $t_{0} = 16$ s 
($\chi^2/d.o.f.=4.02/3)$, marginally consistent with the 
adiabatic evolution (spherical or jet) of $\nu_{\rm m}$.
 
\subsubsection{GRB\ 971208}

GRB971208 is the longest burst ever detected with BATSE. The temporal 
structure of the burst is a simple smooth FRED lasting several thousand 
seconds. The emission is soft, with no emission in channel 4 ($E > 300$ 
keV). The spectral parameters tend to favor fast-cooling, but not 
strongly, as the value of $p=4.06 \pm 0.04$ is unusually high. The value 
of $\Delta = 1.48 \pm 0.04$ is well-determined, and very far from the 
value expected for slow-cooling ($\Delta = 0.5$). Although in apparent 
contradiction to this, the CCD pattern for this event (Figure 7) shows 
a strong resemblance to that of the tail of GRB9890923 in Figure 5.

\subsubsection{GRB\ 980301}

GRB980301 shows a low-energy slope consistent with fast-cooling 
but also a value of $p = 2.48 \pm 0.13$, remarkably consistent with 
values of observed afterglows. The change in slope across the break 
energy is slightly higher than that expected for slow cooling (but 
within two-sigma). If the spectrum is fast-cooling, then we expect 
$\Delta = 0.74 \pm 0.13$, based on the measured high-energy slope. 
This value is within one-sigma of the value $\Delta = 0.69 \pm 0.13$ 
that we derive from the measured slopes. For radiative evolution we 
expect $\beta = -1.84 \pm 0.13$, while for adiabatic evolution we 
expect $\beta = -1.36 \pm 0.13$. However, we measure a much steeper 
value of $\beta = -2.50_{+0.05}^{-1.58}$, suggesting an evolution 
inconsistent with the hydrodynamics of a spherical blast wave, but
consistent with that of a jet. The CCD pattern for GRB980301 is 
shown in Figure 6. Although very similar to the model pattern, 
the observed pattern appears to be displaced.

\subsubsection{GRB\ 981203}

The measured low and high-energy spectral indices for this event 
are notably different than those of other bursts listed in Table 
2. The low-energy spectral index is consistent with the spectral
slope below $\nu_{\rm m}$ in the slow-cooling mode and below 
$\nu_{\rm c}$ in the {\it fast-cooling} regime. Interestingly, for
the {\it fast-cooling} regime the high-energy index is marginally
consistent with the spectral slope for 
$\nu_{\rm c} < \nu < \nu_{\rm m}$. The direct implication here is 
that $\nu_{\rm m}$ is above the BATSE window and has yet to evolve 
through. Hence the value of $p$ is undetermined. The flux in the 
tail was too weak to follow the evolution of the spectrum with any 
reasonable accuracy. From equation 3 and 4, clearly the temporal 
decay should be very shallow, unlike our measured value of 
$\beta = -1.61_{+0.002}^{-0.013}$. This evolution is not consistent 
with the evolution of a spherical blast wave into a constant density 
medium.  

\section{Discussion}

The diverse temporal and spectral properties of GRBs leave their 
origin open to different interpretations. From our analysis, we 
have identified a subset of gamma-ray bursts that exhibit smooth 
high-energy ($\sim$ 25-300 keV) decay emission whose spectral 
properties are very similar to that of fast-cooling synchrotron 
emission that results from a power-law distribution of relativistic 
electrons accelerated in a forward external shock. The 25-300 keV
time histories of the high-energy afterglow candidates are shown 
in Figure 9. The diversity of the time profiles suggests that the
GRB time history is {\it not} necessarily the distinguishing feature
of external shock emission in the fireball model. The diversity
also suggests that the afterglow may be disconnected from the
burst emission (e.g., GRB960530), or overlap the burst emission
(e.g., GRB920622). {\it BeppoSAX} has demonstrated the existence 
of both cases: overlap or continuation of  the afterglow onset 
with the prompt burst emission [e.g., GRB970508 (\cite{piro98}), 
and more recently GRB990510 (\cite{pian01})], and the case in which 
the afterglow begins at a later time, disconnected from the prompt 
GRB [e.g., GRB970228 (\cite{costa00})]. Our analysis further suggests
that in some cases (e.g., GRB971208) the early high-energy afterglow 
may actually be the burst emission itself.  This situation could arise
if the energy deposition in the internal shocks is too low.

\begin{figure*}
   \centerline{\psfig{figure=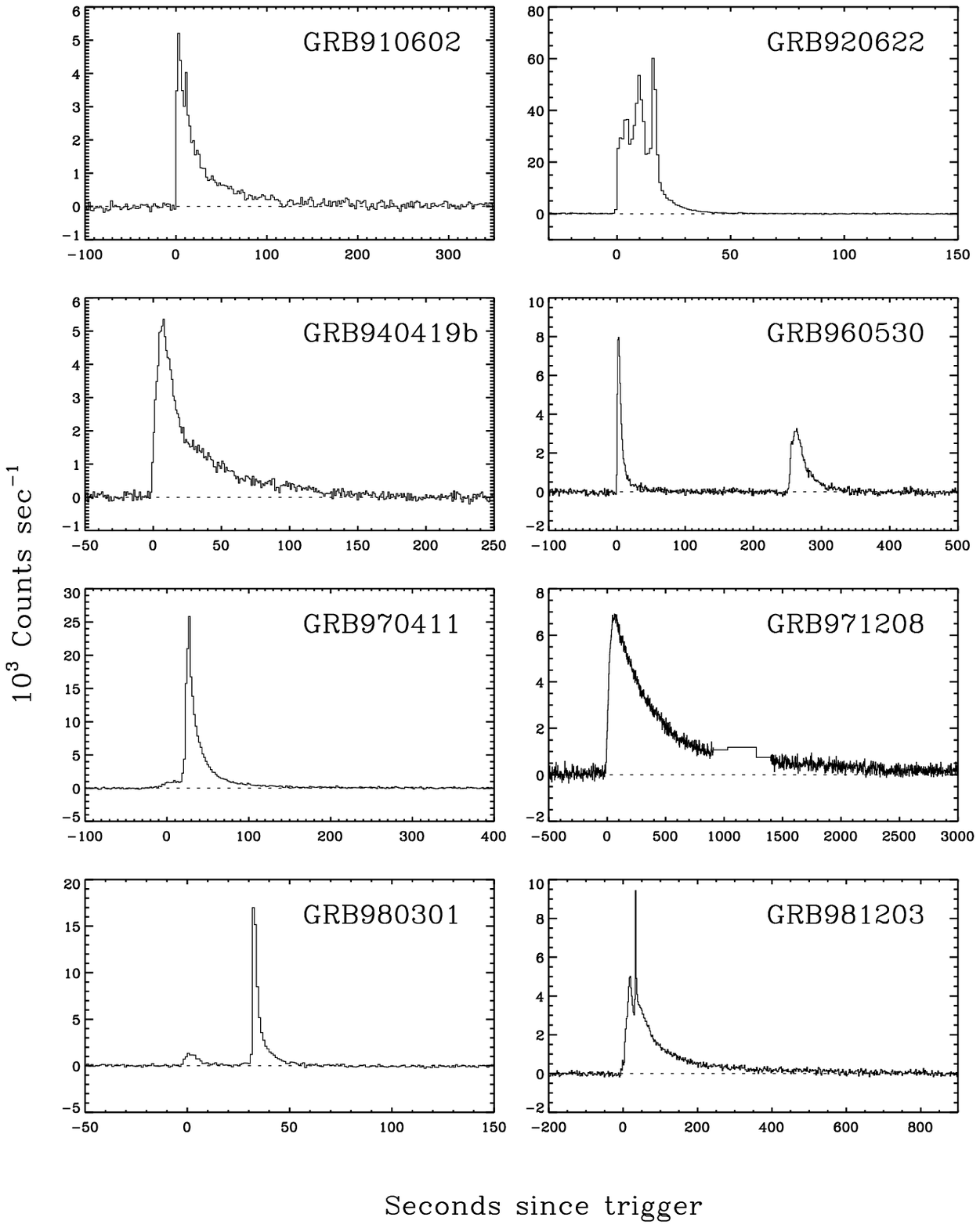,width=7.0truein}}
   \figcaption[fig9.ps]{Time histories (25-300 keV) for the eight 
   high-energy afterglow candidates. \label{figure9}}
\end{figure*}

The spectra of a significant fraction of bursts in our sample, 
however, show inconsistencies with the synchrotron model. From a 
catalog of BATSE GRB spectra, we see that the low and high-energy 
spectral indices follow well-defined distributions (\cite{preece00}). 
For example, the distribution of low-energy power-law indices given 
in Figure 7 of Preece et al.\ (2000) peaks near $\alpha_{\rm low} \sim -1$. 
Roughly $\sim 200$ of the 5500 spectra in the distribution are consistent 
with the expected value of $\alpha_{\rm low}$ below $\nu_{\rm m}$, or about 
$\sim 4\%$. If we adopt the hypothesis that GRB spectra are {\it not} 
synchrotron spectra, then on average $\sim 4\%$ of the time we expect 
to measure parameters consistent with the synchrotron spectrum purely 
by chance coincidence. This implies that we can expect $\sim 0.8$ 
events from Table 2 to have $\alpha_{\rm low} = -1.5$ ($\alpha = -0.5$) 
purely by chance. Clearly, our total of eight candidate events exceeds 
this limit. These events are thus likely sources of synchrotron emission.
\begin{deluxetable}{lccc}
   \footnotesize
   \tablewidth{340pt}
   \tablecaption{Jet Candidates \label{tbl-4}}
   \tablehead{
   \colhead{GRB} & \colhead{Trigger} & \colhead{$p$} & \colhead{$\beta$}
   }
   \startdata
   910602 &257  &$2.00\pm0.01$\tablenotemark{(1)} &$1.74_{-0.11}^{+0.72}$ \\
   920622 &1663 &$2.96\pm0.26$ &$>2.98$ \\
   920813 &1807 &$2.02\pm0.02$ &$2.10_{-0.04}^{+6.05}$ \\
   970302 &6111 &$1.46\pm0.10$ &$1.49_{-0.11}^{+0.93}$ \\
   970411 &6168 &$2.20\pm0.08$ &$2.06_{-0.20}^{+0.21}$ \\ 
   980301 &6621 &$2.48\pm0.13$ &$2.50_{-0.05}^{+1.58}$ \\
   990316 &7475 &$3.04\pm0.09$ &$>2.20$ \\
   990322 &7488 &$0.84\pm0.06$ &$0.87_{-0.01}^{+0.02}$ \\
   990415 &7520 &$2.56\pm0.16$ &$>2.23$ \\
   \cutinhead{Pre-break Jet Candidates}
   910131 &2151 &$1.78\pm0.11$ &$0.71_{-0.11}^{+0.31}$ \\
   940419b &2939 &$3.32\pm0.55$ &$>1.75$ \\
   960530 &5478\tablenotemark{(2)} &$2.96\pm0.22$ &$1.49_{-0.04}^{+0.77}$ \\
   970925 &6397 &$2.56\pm0.13$ &$1.98_{-0.21}^{+0.31}$ \\
   971208 &6526 &$4.06\pm0.04$ &$1.34_{-0.01}^{+0.11}$ \\
   \enddata
   \tablenotetext{(1)}{This value assumes the spectrum is slow-cooling.}
   \tablenotetext{(2)}{First emisison episode of GRB960530.}
\end{deluxetable}

Recent studies on electron acceleration models for ultra-relativistic 
shocks predict values of the electron index in a narrow range 
$2.0 \le p \le 2.5$ (\cite{gallant99}, \cite{gallant00}). Our 
values maintain a significant dispersion under the assumption 
that the observed high-energy afterglow is equivalent to the 
high-energy slope of the fast {\it or} slow-cooling synchrotron 
spectrum (i.e., $\alpha_{\rm high} + 1 = -p/2$). Electron indices
in Table 2 tend to be steepr on average than $p=2.5$. One possible 
alternative for a high value of $p$ ($p \simeq 3$) may be due 
to a shock generated in a decreasing density, $n \propto r^{-2}$, 
external medium that is the result of a massive stellar wind 
(\cite{chevalier98}, \cite{chevalier99}). On the other hand, Sari 
(2000) pointed out that there is no reason why the value of $p$ 
should be different for wind models. Note that several values of
$p$ in Table 3 are {\it below} $p=2$. Hard electron indices ($1<p<2$) 
have recently been reported for the jet model of GRB00301c 
(\cite{panaitescu01}) assuming a broken power-law electron energy
distribution. Similarly, a jet interpretation for GRB010222 would also 
require a flatter electron index, based on analysis of the {\it BeppoSAX} 
data (\cite{intzand01}). However, In'T Zand et al.\ (2001) show that the 
slowing of the ejecta into the non-relativistic regime yields $p=2.2$.
From a theory viewpoint, Malkov (1999) has shown that it is possible 
to obtain a hard electron electron distribution in Fermi acceleration 
models. While the model predictions for the range of electron indices 
appear somewhat uncertain, the {\it observed} dispersion in $p$ values 
may be strongly linked to the accuracy of the fitted value of 
$\alpha_{\rm high}$. Systematic effects may play a role that 
introduces a bias toward steeper values in the estimation of the 
high-energy power-law index. 

Our analysis shows that the tail temporal decays are well-described by 
a power-law with a mean index $\langle \beta \rangle = -2.03 \pm 0.51$. 
While the spectral parameters of approximately $20 \%$ of the decays in 
our sample are in generally good agreement with the synchrotron spectrum, 
the temporal evolution, in general, does not agree well with the evolution 
of a spherical blast wave in a homogeneous medium. There are alternatives 
that might explain this deviation: first, we only considered fully 
radiative or fully adiabatic evolution. More than likely the fireball 
is neither fully radiative nor fully adiabatic throughout its evolution, 
although at the very early stages the evolution may nearly be fully 
radiative while at latter stages the evolution is completely adiabatic. 
B\"{o}ttcher and Dermer (2000) have considered the early afterglow regime 
with the intermediate cases: partially radiative or partially adiabatic 
blast waves. They find that the temporal decay of the spectral flux in 
the fast-cooling regime is a function of 
$\epsilon = \epsilon_{e} \epsilon_{\rm rad}$, where 
$\epsilon_{\rm rad}$ is the fraction of energy radiated by the 
accelerated electrons, or 
$F_{\nu} \propto \nu^{-1/2} t^{-2{(1+\epsilon) \over (8-\epsilon)}}$ 
for $\nu_{\rm c} < \nu < \nu_{\rm m}$ and 
$F_{\nu} \propto \nu^{-p/2} t^{-{{2(1+\epsilon) + 6(p-1)} \over {(8 - 
\epsilon)}}}$ for $\nu > \nu_{\rm m}$. Higher efficiencies therefore 
produce steeper temporal slopes. However, as can be seen for 
$\epsilon = 0.8$, we obtain $F_{\nu} \propto \nu^{-1/2} t^{-1/2}$ 
for $\nu_{\rm c} < \nu < \nu_{\rm m}$. This is steeper than the 
expected value of $-1/4$ from equation 4 but not steep enough to 
match the discrepancies in our observations.

Another possibility to consider is a jet-like geometry or collimated
outflow, as opposed to the simple spherical blast wave. The break in
the light curve may occur at early times after the initial shock, as 
in the case of GRB980519, where evidence exists for a break to a steep 
decay that apparently occurred during the few hours between the GRB and 
the first afterglow detection (\cite{sari99b}). Rhoads (1999) has shown 
for adiabatic evolution that the time of the break in the observer's 
frame goes as $t_{b} \propto \theta_{c}^{2}$. A very early break 
therefore requires a very small $\theta_{c}$. If 
$\theta_{c} < \theta_{b}$ initially, then the slope is steep from the 
start. This implies one of two possiblities: (1) very narrow emission 
spots, or ``nuggets'', within a narrow collimation angle, or (2) a very 
small value of $\Gamma$ such that $\Gamma^{-1} > \theta_{c}$. The second 
option is not likely since the observed emission is in the keV to MeV 
range and $\nu_{\rm m} \propto \Gamma^{4}$, requiring a high Lorentz 
factor. A list of events from Table 2 and 3 with comparable electron 
indices and temporal indices (as required for a jet-like blast-wave) 
are presented in Table 4. Note that although the value of $p$ for 
GRB990322 is low, the temporal decay does follow the $t^{-p}$ relation. 
Events with values of $\beta$ shallower than $p$ may be events in which 
the break occurs at some later time after the GRB. We categorize these 
events as pre-break jet candidates. Three of these events (GRB940419b, 
GRB960530, and GRB970925) have some spectral properties characteristic 
of the synchrotron spectrum (see last column of Table 2). The five 
bursts labeled in Figure 3 are also candidates for jet outflows. More 
importantly, note that three of these events (GRB920622, GRB980301, and 
GRB970411) are strong candidates because they belong to the group of 
high-energy afterglow candidates presented in $\S4$ that were selected 
based on their spectral properties. The spectra of the remaining two 
events (GRB920801 and GRB931223) are only marginally consistent with 
synchrotron emission from an external shock.

\section{Summary and Conclusion}

Our temporal and spectral analysis of the smooth extended gamma-ray decay 
emission in GRBs has shown evidence of signatures for early high-energy 
afterglow emission in gamma-ray bursts. The extended decay emission is 
best described with a power-law function $F_{\nu} \propto t^{\beta}$
rather than an exponential, similar to the results of Ryde and Svensson 
(2001) who studied the decay phase of a sample of GRB pulses with a broad 
range of durations. From our sample of 40 events, we find 
$\langle \beta \rangle \approx -2$ for long, smooth decays. Color-color 
diagrams have provided a qualitative interpretation of the burst spectral 
evolution and allow a simple comparison with the evolution expected from 
the synchrotron model as well as comparison of spectral evolution among 
GRBs. The CCD patterns and the spectral analysis indicate that $\sim 20\%$ 
of the events in our sample are consistent with synchrotron emission expected 
from an external shock. Interestingly, three of these events have decay rates 
consistent with that expected from the evolution of a jet, 
$F_{\nu} \sim t^{-p}$. Because the break is essentially at the onset of 
deceleration, the jet must, at least, be very narrow, since 
$\theta_{c} < 1/\Gamma$. Table 4 suggests that in some cases the break 
occurs at a later time, so that the prompt emission we observe is pre-break, 
$\theta_{c} > 1 / \Gamma$ and consistent with spherical geometry. A possible 
scenario is one in which the ejecta is very grainy, where the nuggets in the 
ejecta are smaller than $1/\Gamma$, similar to the model discussed by Heinz 
and Begelman (1999). Huang et al.\ (1999) (see also Wei and Lu 2000) have 
shown that the break in the light curve is more of a smooth transition due 
to the off-axis emission of a jet with no angular dependence. The steep 
light curve can only occur if the angular size of the nugget is less than 
$1/\Gamma$.

Connaughton (2001) has investigated the average late-time temporal 
properties of GRBs observed with BATSE and found statistically 
significant late time power-law decay emission that softens relative 
to the initial burst emission, suggesting the existence of early 
high-energy afterglow. Other studies using PHEBUS (\cite{tkachenko00})
and APEX (\cite{litvine00}) bursts show similar behaviors in late-time
GRB light curves. Collectively, these studies strongly suggest that the 
afterglow emission may overlap or be connected to the prompt, variable 
burst emission. On the other hand, it is clear that not all GRBs exhibit 
such behavior. In some cases, the initial gamma-ray flux from the external
shock may simply be too low to detect (e.g., see Figure 4 in Giblin et al.\
2000). In other cases, the bulk Lorentz factor may be too low to generate
the gamma-ray photons upon impact with the surrounding medium. 

As the number of afterglow/counterpart detections increases, the 
relationship of the afterglow emission to the gamma-rays released 
in the initial phase of the burst can be studied systematically. 
The capabilities of {\it Swift} (\cite{gehrels02}) will allow broad 
spectral coverage using three co-aligned instruments (BAT, XRT, and 
UVOT) during the gamma-ray phase and early afterglow phase of the 
burst and facilitate the distinction between the GRB and the onset 
of the afterglow based on temporal {\it and} spectral information.  
With well-constrained spectral and temporal parameters in hand, plots 
of temporal index vs.\ spectral index can be readily constructed and 
thus provide information on the geometry of the fireball and definitively 
test the internal/external shock model for GRBs.

\acknowledgements

We thank the referee for valuable comments that enhanced this paper. 
We are also grateful to Jon Hakkila and the late Robert Mallozzi for 
numerous helpful discussions. Tim Giblin, Ralph Wijers, and Valerie 
Connaughton acknowledge support from NAG5-11017.  During preparation 
of this manuscript, the BATSE Team and GRB community lost two outstanding 
members:  Jan van Paradijs, and Robert Mallozzi. Jan continued to work on 
this project until the few days before his death. Their contributions to 
this work and our knowledge of GRBs will not be forgotten. Moreover, they 
will be sorely missed as both our colleagues and friends.

\end{document}